\documentclass{ws-ijmpa_1}

\newcommand {\half} {{1 \over 2}}
\newcommand {\hhalf} {{\textstyle{1 \over 2}}}

\newcommand {\bra}[1] {\left< #1 \right|}
\newcommand {\ket}[1] {\left| #1 \right>}
\newcommand {\expect}[1] {\left< #1 \right>}
\newcommand {\braket}[2] {\left< { \left. #1 \,\right|} \,#2 \right>}
\newcommand {\Tr} {\mathrm{Tr}\,}

\begin{document}

\title{
       Coordinate-invariant Path Integral Methods in Conformal Field Theory}

    \author{Andr\'e van Tonder}

\address{   Department of Physics, Brown University, Box 1843 \\
            Providence, RI 02912, U.S. \\
            andre@het.brown.edu}

    \maketitle

   \begin{history}
    \received{21 February 2005}
    \
\end{history}

    \begin{abstract}
        
We present a coordinate-invariant approach, based on a Pauli-Villars
measure, to the definition of the path 
integral  in two-dimensional 
conformal field theory.  We discuss some advantages of this
 approach compared to the operator formalism and 
alternative path integral approaches.  We show that our path integral
measure is invariant under conformal transformations and field
reparametrizations, in contrast to
 the measure used 
in the Fujikawa calculation, and we show the agreement,
 despite different 
origins, of the conformal anomaly in the two approaches.
The natural  
energy-momentum in the Pauli-Villars approach 
is a true coordinate-invariant tensor quantity, 
and we discuss its nontrivial relationship to the corresponding non-tensor 
object arising in the operator
formalism, thus providing
a novel explanation within a path integral context
for the anomalous Ward identities 
of the latter. 
We provide a direct  
calculation of the nontrivial contact terms arising
in expectation values of certain energy-momentum products, 
and we use these to perform a simple consistency check 
confirming the validity of the change of
variables formula for the path integral.  
Finally, we review the 
relationship between the conformal anomaly and the energy-momentum 
two-point functions in our formalism.   


\end{abstract}

        \ccode{PACS numbers:
          11.25.Hf, 11.10.Gh}

\section{Introduction}

In the conformal field theory literature, path integrals are often 
used for heuristic derivations.  Operator methods or 
axiomatics are then invoked to give operational meaning to the 
resulting expressions.  

Doing this is highly dangerous.  It is easy to overlook the fact
that identities derived using path integral methods may be
invalidated by replacing insertions by the obvious operator
quantities.  In fact, the relationship between natural path integral
and natural operator quantities can be highly nontrivial.
Some lack of appreciation  of this issue in the literature
 clearly calls for a more 
careful comparative analysis.    

There are few direct path integral calculations in the conformal
field theory literature.  Apart from some derivations 
of the conformal anomaly, done for example using the Fujikawa 
approach, careful path integral calculations of many nontrivial 
results, including Ward identities,
seem to be absent.  Here we provide a few
such calculations.

A disadvantage
of the operator formalism is that the definitions of 
the operators typically depend on a choice of coordinates, due to 
the chosen regularizations.  The coordinate-invariance of the 
original Lagrangian is neither respected nor exploited. 
Changes of coordinates are conflated with implicit anomalous gauge 
transformations, complicating the transformation laws and 
obscuring the physical origin of the anomalies.  

Some common approaches to path-integrals suffer from 
comparable problems.  Dimensional, zeta-function and and heat kernel 
regularization techniques
all introduce implicit dependencies on extra structure on the world
sheet via  
choices of coordinates, cutoff, complex structure,
or through the lack of a canonical way of regulating interesting 
operator insertions.   
In addition, neither dimensional 
nor zeta-function techniques are
adaptable to a non-perturbative definition of the path 
integral measure.

The Fujikawa approach provides a way of defining 
a regulated path integral measure, but does not come with a 
corresponding prescription for regulating interesting operator insertions.
It is therefore incomplete.    

In this paper we address all these issues by
presenting a coordinate-invariant approach, based on a
 Pauli-Villars regularization, to the definition
of the path integral measure 
and the calculation of anomalies in a two-dimensional scalar
conformal field theory.  
The measure does not depend on a choice of coordinates or complex 
structure on the world sheet
and is largely insensitive to the details of the cutoff procedure.
In contrast to the Fujikawa measure, it is invariant under conformal 
transformations and field reparametrizations, and it 
simultaneously regulates also the interesting operator insertions.
In contrast to dimensional or zeta-function
techniques, the Pauli-Villars measure is suitable for a 
nonperturbative definition of the path integral.
In contrast to the operator approach, the natural insertions are 
regulated in a coordinate-invariant way, and are true tensors.

By comparing the path integral 
measures in the two cases, we show the consistency, despite apparently different 
origins, of the conformal anomaly in the Pauli-Villars 
and the Fujikawa approaches. 
In the Fujikawa approach, the anomaly arises as a dependence
of the path integral measure on the background metric.  On the other 
hand, we show how the full
Pauli-Villars path integral measure is invariant under 
conformal transformations of the background, 
and the anomaly is shifted to the 
expectation value of the trace of the energy-momentum tensor, which 
becomes nonzero. 
 While the conformal anomaly has been 
derived before using Pauli-Villars regulators, previous work has either 
been silent on the proper definition of the measure or 
introduced new, nonstandard measures for the Pauli-Villars auxiliary fields.
Here we only use ordinary bosonic and Grassmann auxiliary fields.
  
We point out that there is a nontrivial relationship
between the energy-momentum tensor 
in the path-integral formalism and the corresponding 
quantity in the operator formalism.
We show that the natural  
energy-momentum in the Pauli-Villars approach 
is a true coordinate-invariant tensor quantity, 
which we then relate to the more familiar non-tensor 
object arising in the operator
approach, obtaining
a new explanation within the context of the path integral 
for the anomalous properties 
of the latter. 

We confirm by explicit calculation that the full energy-momentum tensor
satisfies \textit{classical\/} Ward identities, and 
we explain how these are related to the anomalous Ward identities found
in the operator formalism.  

We provide a first direct path-integral 
calculation of the nontrivial contact terms arising
in expectation values of certain energy-momentum products, 
previously derived only using axiomatic considerations, and
we use these to perform a simple consistency check confirming the change of
variables formula for the Pauli-Villars path integral.
Finally, we review the 
relationship between the conformal anomaly and the energy-momentum 
two-point functions in our formalism.  Here the contact terms 
are essential to obtaining the correct results.

\section{The Fujikawa approach}

In this section we derive the conformal anomaly using an adaptation
of the methods of Fujikawa,\cite{fujikawa1,fujikawa2,nakahara}
according to which anomalies arise
as a non-invariance of the path integral measure 
under classical symmetries.  We present the full 
calculation, both because our method is simpler than existing
presentations, and because we will reuse aspects of it 
in our calculation of the transformation of the Pauli-Villars measure
in the next section.  There we will provide 
a derivation of the anomaly using a Pauli-Villars definition of
the path integral measure, showing that the origin of the anomaly
is shifted from the measure to the expectation value of the trace of the 
energy-momentum tensor.  

We are interested in the dependence of the partition function
$$
  Z(g) \equiv \int [d\phi]_g\, e^{-S(g, \phi)}
$$
on the background metric $g_{ij}$.  Here $[d\phi]_g$ denotes the
$\infty$-dimensional differential form
$$
  [d\phi]_g \equiv \bigwedge_{n = 0}^\infty da_n^g = da_0^g \wedge
  da_1^g \wedge \cdots,
$$
where the coefficients $a_n^g:\phi\to \mathbf{R}$ depend on the 
metric via the expansion
$$
  \phi(z) = \sum_{n=0}^\infty a_n^g\,\phi_n^g(z),
$$
and where $\phi_n^g(z)$ denotes an orthonormal basis of field configurations
satisfying
\begin{align}
  \int d^2x \sqrt{g}\,\phi^g_m(x)\,\phi^g_n(x) = \delta_{mn}. \label{ortho}
\end{align}
To calculate the variation $\delta_g[d\phi]_g$ of the path integral 
measure with respect to deformations of the metric, we will need the
variation
\begin{align}
  \delta_g a^g_m &= \delta_g \int d^2x\,\sqrt{g}\,\phi^g_m\phi \\
    &= \int d^2x\,\sqrt{g}\left(\half g^{ij}\delta g_{ij}\right)
           \phi^g_m\phi
    + \int d^2x\,\sqrt{g}\,
           (\delta\phi^g_m)\phi.   \label{variation}
\end{align}
The metric dependence of $\phi^g_m$ is not uniquely fixed by 
the above orthonormality requirement.  However, given a metric
$g$, any two orthonormal bases are related by a unitary transformation
that will leave the form $[d\phi]_g$ invariant.  Therefore we can,
without loss of generality, choose one particular metric dependence 
for $\phi^g_m$ compatible with orthonormality.  To find such a choice,
we 
vary both sides of (\ref{ortho}) with respect to the metric, obtaining
\begin{align*}
  0 = \int d^2x\,\sqrt{g} \left(\half\,g^{ij}\delta g_{ij}\right)
         \phi_m^g\phi_n^g 
      + \int d^2x \,\sqrt{g}\,(\delta_g\phi_m^g)\,\phi_n^g
     + \int d^2x \,\sqrt{g}\,\phi_m^g\,(\delta_g\phi_n^g).
\end{align*}
A suitable choice for $\delta_g\phi_m^g$ is therefore
$$
  \delta_g\phi_m^g \equiv -{1\over 4}\,g^{ij}\delta g_{ij}\,\phi_m^g.
$$ 
Inserting this in (\ref{variation}) gives
\begin{align*}
   \delta_g a^g_m 
    &= \int d^2x\,\sqrt{g}\left({1\over 4}\, g^{ij}\delta g_{ij}\right)
           \phi^g_m\phi   \\
    &= \sum_n a^g_n\cdot 
       \int d^2x\,\sqrt{g}\left({1\over 4}\, g^{ij}\delta g_{ij}\right)
           \phi^g_m\phi^g_n  \\
   &\equiv \sum_n a^g_n \,C_{mn},
\end{align*}
where
$$
   C_{mn} \equiv \int d^2x\,\sqrt{g}\,\phi^g_m\left({1\over 4}\, g^{ij}\delta g_{ij}\right)
           \phi^g_n
$$
We then find, using the normal rules for manipulating differential forms, 
\begin{align}
  \delta_g[d\phi]_g &\equiv \delta_g \left(\bigwedge_n da^g_n\right) \nonumber\\
   &= \sum_m \cdots\wedge da^g_{m-1}\wedge\delta(da^g_m)\wedge
         da^g_{m+1} \wedge \cdots \nonumber\\
   &=  \sum_m \cdots\wedge da^g_{m-1}\wedge\left(\sum_n da^g_n\,C_{nm}\right)\wedge
         da^g_{m+1} \wedge \cdots \nonumber \\
   &= \left(\sum_m C_{mm}\right)\bigwedge_n da^g_n \nonumber \\
  &= \left(\Tr C\right)_g\, [d\phi]_g,    \label{trace}
\end{align}
where $C$ is the operator
$$
  C \equiv {1\over 4}\, g^{ij}\delta g_{ij} = 
    \half \,{1\over\sqrt{g}}\,\delta\sqrt{g} \equiv \delta\omega.
$$
Notice that $\delta\sqrt{g}$ is the local change of volume.  In other
words, the path integral measure will be scale dependent.  This is the
origin of the conformal anomaly.  

We now need to calculate the trace
$$  
  \left(\Tr C\right)_g = \int d^2x\,\sqrt{g}\,\delta\omega(x)\, A(x),
$$  
where the infinite sum
$$
  A(x) \equiv \sum_m \phi^g_m(x)\,\phi^g_m(x)
$$
does not in general converge.  A natural short-distance 
regularization, which can be taken as part of the definition of the  
path integral, is obtained
by considering instead the limit as $\epsilon \to 0$ of the 
sum\cite{difrancesco,albeverio}
$$
  \sum_m \phi^g_m(x)\, e^{\epsilon \Delta}\,\phi^g_m(x)
    = \left<x |e^{\epsilon \Delta}|x\right>
$$
where $\Delta$ denotes the Laplacian
$$
  \Delta \equiv {1\over \sqrt{g}}\,\partial_i\left[\sqrt{g}\,g^{ij}\partial_j
     \,(\, \cdot\,) \right].
$$
Here the position basis bras and kets are defined via 
$\braket{x}{y} \equiv \delta(x-y) / \sqrt{g}$, to make
$\braket{x}{\phi} = \phi(x)$ with respect to the inner product 
$\braket{\phi_1}{\phi_2} = \int \sqrt{g} \,\phi_1\,\phi_2$.  

As required for consistency with our previous remark that a unitary
transformation leave $[d\phi]_g$ invariant, 
the regularized expression $\left<x |e^{\epsilon \Delta}|x\right>$
depends only on the metric and not on the particular 
orthonormal basis  $\phi_m^g$ that we have chosen.  Also 
notice that if we insert an eigenbasis $\ket{m}$ of $\Delta$, the above becomes 
$$
\left<x |e^{\epsilon \Delta}|x\right> = 
  \sum_m \braket{m}{x} \braket{x}{m}\, e^{\epsilon \,\lambda_m}
$$
where the $\lambda_m$ are the corresponding (negative) eigenvalues, 
making it clear that
the regularization corresponds to a large-momentum suppression.  

We therefore need to calculate the small $t$ behaviour of the 
function
$$
  G(x,t|y,0) \equiv \theta(t)\bra{x} e^{t\Delta}\ket{y},
$$
where $\theta(t)$ denotes the step function, inserted to avoid the
regime $t<0$ where $\bra{x} e^{t\Delta}\ket{y}$ diverges.  
It is easy to check
that
$$
  \left(\partial_t - \Delta_x\right) \, G(x,t|y,0) = \delta(t)\braket{x}{y}
     = {1\over \sqrt{g}}\, \delta(t)\,\delta(x - y).
$$    
In other words, the distribution $G(x,t|y,0)$ solves the heat or diffusion 
equation given a point source at $y$ at time $t = 0$,  
and is known as a \textit{heat kernel}.  
Without loss of generality, we can choose coordinates so that $y = 0$.     
In two dimensions, we can further choose the coordinate system so that
\begin{align*}
  g_{ij} (x) &= e^{2\omega}\,\delta_{ij}, \\
  \omega (0) &= 0.
\end{align*}
so that the equation for $G(x,t|y,0)$ becomes
\begin{align*}
  (\partial_t - e^{-2\omega (x)}\Delta_0)\, G(x,t|0,0) &= \delta(t)\,\delta(x),
\end{align*}
where $\Delta_0 \equiv \delta^{ij}\partial_i\partial_j$.  Regarding this
as an operator equation in the space $\mathcal{L}^2 (\mathbf{R}^2)$, we
 write
the solution as 
$$
  G(x,t|0,0) = \bra{x,t} \left({1\over \partial_t - e^{-2\omega (x)}\Delta_0}\right)\ket{0,0}.
$$
Since we 
are interested in the limit as $t\to 0$, for which the diffusion 
becomes increasingly short-ranged, 
we will
develop an expansion for $G(0,t|0,0)$ in terms of derivatives of 
$\omega$ at the origin.  
Expanding
around $x^i = 0$, we have 
\begin{align*}
  e^{-2\omega(x)} = 1 - 2\,(\partial_i \omega)\, x^i
    + \left[-(\partial_i \partial_j\omega) +
    2 \,(\partial_i \omega)\,(\partial_j \omega)\right]  x^i x^j  + \cdots,
\end{align*}
where the derivatives are all evaluated at $x^i = 0$.  We can then 
write\cite{difrancesco}
$$
  G = {1\over A - B} = {1\over A} +  {1\over A}\,B\, {1\over A}
     +  {1\over A}\,B\, {1\over A}\,B\, {1\over A} + \cdots
$$
where $A \equiv \partial_t - \Delta_0$, so that $1/A$ is the flat space solution
$$
  \bra{x, t} \left({1\over A}\right)\ket{0, 0}
    = {1\over 4\pi t}\, e^{-x^2 / 4 t}\, \theta(t)
$$
while 
$$
  B \equiv \left \{- 2\,(\partial_i \omega)\, x^i
    + \left[-(\partial_i \partial_j\omega) +
    2 \,(\partial_i \omega)\,(\partial_j \omega)\right] x^i x^j + \cdots
   \right\} 
    \Delta_0
$$
Inserting this expansion in the first-order contribution 
$A^{-1} BA^{-1}$ to $G(0,\epsilon|0,0)$, 
the first term, with odd 
integrand proportional to $x^i$, vanishes.  The next term in $A^{-1} BA^{-1}$
is proportional to 
\begin{align*}
  \int_0^\epsilon dt &\int d^2x\, {1\over 4\pi\,(\epsilon - t)}\,
       e^{-x^2/4\,(\epsilon - t)} \, x^i x^j\,\Delta_0\,{1\over 4\pi t}
           \,e^{-x^2/4t} \\
 &= \int_0^\epsilon dt\int d^2x\, {1\over 4\pi\,(\epsilon - t)}\,{1\over 4\pi t}
      \, e^{-x^2/4\,(\epsilon - t)} \, x^i x^j
        \left[-{1\over t} + {x^2\over 4 t^2} \right]
           \,e^{-x^2/4t} \\
&= \half \,\delta^{ij}\int_0^\epsilon dt\int d^2x\, {1\over 4\pi\,(\epsilon - t)}\,{1\over 4\pi t}
      \, e^{-x^2/4\,(\epsilon - t)} \, x^2
        \left[-{1\over t} + {x^2\over 4 t^2} \right]
           \,e^{-x^2/4t} \\
&= \half \,\delta^{ij}\int_0^\epsilon dt\int d^2x\, {1\over 4\pi\,(\epsilon - t)}\,{1\over 4\pi t}
      \, x^2
        \left[-{1\over t} + {x^2\over 4 t^2} \right]
           \,e^{-\epsilon\,x^2 / 4t\,(\epsilon-t)}.
\end{align*}
Writing the terms containing $x^2$ and $x^4$ as derivatives with respect
to $\lambda \equiv \epsilon/ 4t\,(\epsilon-t)$ of 
the Gaussian integral 
$$
  \int d^2x\, e^{-\lambda x^2} = {\pi\over \lambda},
$$
this simplifies to 
\begin{align*}
  {1\over 2\pi}\,\delta^{ij} \int_0^\epsilon dt\,
    \left\{- {(\epsilon - t)\over \epsilon^2} 
      + {2\,(\epsilon - t)^2 \over \epsilon^3}\right\}
   &= {1\over 12 \pi}\, \delta^{ij}
\end{align*}
The next nonzero term in the contribution $A^{-1} BA^{-1}$
has integrand proportional to $x^ix^jx^kx^l$.  This may be checked to be
of order $\epsilon$.  Further terms are of even higher order in $\epsilon$, 
so that the contribution to $G$ from the term  $A^{-1} BA^{-1}$ can be written 
$$
  G_1(0,\epsilon|0,0) = {1\over 12\pi}\, 
     \left[-\Delta\omega +
    2 \,\partial_i \omega\partial^i \omega\right] + o\left({\epsilon}\right).
$$

There is one additional contribution of order $\epsilon^0$ to $G(0,\epsilon|0,0)$.  
It comes
from the second-order term $A^{-1}BA^{-1}BA^{-1}$ in the above expansion.
It is 
\begin{align*}
   &4\,(\partial_i \omega)\,(\partial_j \omega)
   \int_0^\epsilon dt \int_0^t du
    \int d^2x \int d^2y \, \times \\
    &\quad \times
       {1\over 4\pi(\epsilon - t)}\,e^{-x^2/4(\epsilon- t)}
       \,x^i\,\Delta_0^x\, {1\over 4\pi(t-u)}\,e^{-(x-y)^2/4(t - u)}\,
         y^j\,\Delta_0^y\, {1\over 4\pi u}\,e^{-y^2/4u}
\end{align*}
By similar manipulations, this becomes
\begin{align*}
    -{1\over 6\pi} \,\partial_i \omega\partial^i \omega, 
\end{align*}
so that 
\begin{align*}
  G_2(0,\epsilon|0,0) =  -{1\over 6\pi} \,\partial_i \omega\partial^i \omega
    + o\left(\epsilon\right)
\end{align*}
Notice that this contributes a term that exactly cancels the
term of the same form in $ G^2(0,\epsilon|0,0) $.  
We find the result
\begin{align*}
  G (0,\epsilon|0,0) &= {1\over 4\pi\epsilon} 
       -
       {1\over 12\pi}\,  
                  \Delta\omega +  o\left(\epsilon\right) \\
  &= {1\over 4\pi\epsilon} 
     +  {1\over 24\pi}\, 
                  R +  o\left(\epsilon\right),
\end{align*}
where we have used 
$$
  R = -2\,e^{-2\omega} \, \Delta \omega
$$
and $\omega (0) = 0$.  
Inserting our result into the formula for the variation of the
measure, we obtain
$$
  \delta_g\,[d\phi]_g =    {1\over 4\pi\epsilon}\int d^2x\,\sqrt{g}\,\delta\omega(x) + {1\over 24\pi} \int d^2x\,\sqrt{g}\,\delta\omega(x)\,R.
$$
The first term diverges as $\epsilon\to 0$, but may be exactly canceled by adding
the counterterm 
$$
   {1\over 8\pi\epsilon}\int d^2x\,\sqrt{g}
$$
to the original action.  After doing this, we obtain our final result
\begin{align}
  \delta_g\int [d\phi]_g\, e^{-S(\phi, g)} = \left( {1\over 24\pi} \int
  d^2x\,\sqrt{g}\,\delta\omega(x)\,R \right) \int [d\phi]_g\, e^{-S(\phi, g)}.
\label{confanom}
\end{align}

\section{A Pauli-Villars derivation of the anomaly}
\label{PV}

In the previous section the conformal anomaly was obtained from 
the dependence of the path integral measure on the metric.  
In the case of a Weyl transformation 
$g_{ij} \to e^{2\omega}  g_{ij}$, we could absorb the anomalous dependence 
of the measure into the energy-momentum tensor by defining a modified 
energy-momentum tensor $\tilde T_{ij}$ via
\begin{align*}
  \delta \int [d\phi]_g\,e^{-S(g, \phi)}
    &\equiv {1\over 4\pi} \int d^2x\,\sqrt{g}\,\delta g^{ij}\expect{\tilde T_{ij}}_g \\
    &= {1\over 4\pi}\int d^2x\,\sqrt{g}\,(-2\,\delta\omega)\,\expect{\tilde T^{\,i}_{\phantom{i}i}}_g.
\end{align*}
Given the action
\begin{align*}
  S(\phi) & = \half\int d^2x \, \sqrt g\, g^{ij}\, \partial_i \phi\,\partial_j\phi,
\end{align*} 
we have
$$
  T_{ij} = -2\pi\left(\partial_i\phi\,\partial_j\phi
             - \half\,
          g_{ij}g^{kl}\,\partial_k\phi\,\partial_l\phi\right).
$$
Comparing with (\ref{confanom}), this would require that we define 
\begin{align}
  \tilde T^{\,i}_{\phantom{i}i} \equiv T^{\,i}_{\phantom{i}i} - {1\over 12}\, R = -{1\over 12}\, R,
\label{tildeT}
\end{align}
since $T^{\,i}_{\phantom{i}i}$ is identically zero.  The curvature term was 
entirely due to the variation $\delta \,[d\phi]_g$ of the integration
measure.    

Note, however, that the Fujikawa method is incomplete, since it does
not provide a canonical choice for regulating $T_{ij}$.  Without 
specifying such a choice, we cannot assume, as we have above,
that $\expect{T^{\,i}_{\phantom{i}i}}$
will indeed be zero, or even finite. 

It is therefore very instructive to derive the conformal anomaly using a Pauli-Villars 
regularization,\cite{PV}  which provides a complete, coordinate-invariant
way of regulating both the measure and the energy-momentum tensor. 
This was first done by Vilenkin in Ref.~\refcite{vilenkin}, though only
up to first order in the expansion of the curvature around flat space, and
ignoring possible nontrivial transformations of the path integral measure.  
For 
related work, including comparisons of the Pauli-Villars method
to the $\zeta$-function and point-splitting methods, see for example 
Refs.~\refcite{birrell}--\refcite{anselmi}.  
These works are either 
silent on the proper definition of the measure or 
introduce new, nonstandard measures for the Pauli-Villars auxiliary 
fields.\cite{diaz,anselmi}
In the following we carefully define the measure using only 
ordinary bosonic and Grassmann auxiliary fields.

With a Pauli-Villars definition of the path integral, 
we shall see that the integration measure will in fact 
be invariant under variations of $g$,\cite{diaz} 
and the curvature anomaly will 
instead be due to the fact that the Pauli-Villars fields are massive, so that
$T^{\,i}_{\phantom{i}i} \ne 0$, and we shall indeed find  that 
\begin{equation}
  \expect{T^{\,i}_{\phantom{i}i}}_g \to  -{1\over 12}\, R
\label{PVanom}
\end{equation}
in the limit where the Pauli-Villars masses go to infinity.

In the following discussions, we will have need to carefully regulate 
the theory in both the infrared and ultraviolet.  For our infrared
regularization we give a small mass $m$ to the field $\phi$.  The ultraviolet
regularization will consist in 
adding auxiliary Pauli-Villars fields that are either real scalars $\chi_m$
with bosonic statistics and action
\begin{align*}
  S(\chi_m) & = \half\int d^2x \, 
          \sqrt g\, \left( g^{ij} \partial_i \chi_m\,\partial_j\chi_m + M_m^2\,
              \chi_m^2\right)
\end{align*} 
or complex scalars $\chi_m$, $\bar \chi_m$ with Grassmann statistics and 
action
\begin{align*}
  S(\chi_m, \bar\chi_m) & = \half\int d^2x \, 
          \sqrt g\, \left( g^{ij} \partial_i \bar\chi_m\,\partial_j\chi_m + M_m^2\,
              \bar\chi_m \chi_m\right).
\end{align*} 
Denoting 
\begin{align}
\bar\chi_m = \chi_m
\label{barconvention}
\end{align} 
for the bosonic fields, we 
can write the Pauli-Villars action in unified form
\begin{align*}
  S_{\textit{PV}} & = \sum_m \half\int d^2x \, 
          \sqrt g\, \left( g^{ij} \partial_i \bar\chi_m\,\partial_j\chi_m + M_m^2\,
              \bar\chi_m \chi_m\right).
\end{align*} 
The masses $M_m$ and statistics of the fields will be 
chosen to obtain finite expectation values for interesting quantities.
In the end, we will be interested in the limits 
$m\to 0$ where the field $\phi$ becomes massless, and $M_m \to \infty$
where the Pauli-Villars fields become non-dynamical.

We first calculate the variation 
$$
\delta_g \biggl([d\phi]_g^{\mathit{PV}}\biggr) \equiv
 \delta_g \biggl([d\phi]_g\wedge[d\chi_1]_g\wedge\cdots\wedge[d\chi_n]_g
 \biggr)
$$  
of the full path integral measure $[d\phi]_g^{\mathit{PV}}$
including  matter and auxiliary fields, where for a bosonic field
$\chi$ the form $[d\chi]_g$ is defined just like $[d\phi]_g$ 
in the previous section, while for a Grassmann field $\chi$, $\bar\chi$,
we define 
$$
  [d\chi]_g \equiv d\chi^g_0\wedge d\bar\chi^g_0 \wedge
           d\chi^g_1\wedge d\bar\chi^g_1 \wedge \cdots.
$$ 
Referring back to our calculation (\ref{trace}),
each real scalar field $\chi_i$ will contribute
a term 
$$\left(\Tr C\right)_g\, [d\chi_i]_g$$
 to the variation.
The contribution of each Grassmann field $\chi_i$ will 
be 
$$- 2 \left(\Tr C\right)_g\, [d\chi_i]_g.$$  
Indeed, the reader may easily check that
with Grassmann integration rules, if $\delta\chi = \epsilon \,\chi$ 
where $\epsilon$ is
a real parameter, then for the 
change of variables formula
$\int d\chi'\, f(\chi') = \int d\chi\, f(\chi)$ to hold (equivalent to 
$\delta \int d\chi\, f(\chi) = 0$) 
one needs $\delta (d\chi) \equiv - \epsilon\, d\chi$.  
Doing this for each factor in $d\chi^g_n\wedge d\bar\chi^g_n$, we find 
a factor of $-2$ relative to the result (\ref{trace}) of the previous 
section.  
  
Defining $c_i = 1$ for $\chi_i$ bosonic 
 and $c_i = -2$ for $\chi_i$ Grassmann, we find 
\begin{align*}
 & \delta_g \biggl([d\phi]_g\wedge[d\chi_1]_g\wedge\cdots\wedge[d\chi_n]_g
 \biggr) \\
  &\qquad\qquad
   =
   \left(1 + \sum_i c_i\right) 
    \left({1\over 4\pi\epsilon}\int d^2x\,\sqrt{g}\,\delta\omega(x) +
 {1\over 24\pi} \int d^2x\,\sqrt{g}\,\delta\omega(x)\,R\right)\times
 \\
&\qquad\qquad\qquad\qquad\qquad\qquad
  \times
 \biggl([d\phi]_g\wedge[d\chi_1]_g\wedge\cdots\wedge[d\chi_n]_g
 \biggr)
\end{align*}
 We therefore can choose our 
Pauli-Villars field statistics to cancel the variation of $[d\phi]_g$.
 In particular, if we choose
\begin{align}
1 + \sum_i c_i = 0,
\label{PV1}
\end{align}
the variation of the full measure is zero
$$
\delta_g \biggl([d\phi]_g^{\mathit{PV}}\biggr) \equiv
 \delta_g \biggl([d\phi]_g\wedge[d\chi_1]_g\wedge\cdots\wedge[d\chi_n]_g
    \biggr) = 0.
$$  
Since the measure is now invariant, the anomaly will have to be due entirely to
the expectation value of the trace of the energy-momentum tensor, i.e.,
\begin{align}
  \delta \int [d\phi]_g^{PV}\,e^{-S(g, \phi, \bar\chi_i, \chi_i)}
    &= {1\over 4\pi}\int d^2x\,\sqrt{g}\,(-2\,\delta\omega)\,\expect{ T^{\,i}_{\phantom{i}i}}_g.
\label{PVvar}
\end{align}
where,
with our convention (\ref{barconvention}),
$$
  \expect{T^{\,i}_{\phantom{i}i}}_g = 2\pi\,\expect{m^2\phi^2
    + \sum_i M_i^2 \bar\chi_i\chi_i}_g,
$$
and the expectation values are finite due to conditions on the 
Pauli-Villars masses to be specified below.  

The propagator of a complex scalar is double that of a real scalar, while
each Grassmann loop takes an additional factor $-1$.  
We therefore find on the plane, with $g_{ij} = \delta_{ij}$,
\begin{align}
  \expect{T^{\,i}_{\phantom{i}i}}_\delta = 2\pi \int {d^2p\over(2\pi)^2} \, \left\{
      {m^2\over p^2 + m^2} + \sum_i c_i  \,{M_i^2\over p^2 +
      M_i^2}\right\}.
\label{flat}
\end{align}
As long
as we impose the condition
\begin{align}
  m^2 + \sum c_i M_i^2 = 0,
\label{PV2}
\end{align}
on the Pauli-Villars masses, the integral converges.   The answer in this case is
\begin{align*}
  2\pi\cdot{1\over 4\pi} &\lim_{\Lambda\to\infty} \left(m^2 \ln {\Lambda^2\over m^2} + 
       \sum_i c_i\, M_i^2 \ln {\Lambda^2\over M_i^2}\right) \\
   &= -\half \left(m^2 \ln {m^2\over \mu^2} + 
       \sum_i c_i\, M_i^2 \ln {M_i^2\over \mu^2}\right),
\end{align*}
where the limit is finite due to the Pauli-Villars condition (\ref{PV2}).  
Here 
$\mu$ is an arbitrary parameter with dimension of mass, irrelevant 
due to the condition (\ref{PV1}).  As $m\to 0$,
the first term 
in (\ref{flat}) falls
away and we get  
\begin{align*}
   -\half
       \sum_i c_i\, M_i^2 \ln {M_i^2\over \mu^2},
\end{align*}
This diverges in the limit $M_i\to\infty$ but can be exactly compensated by 
a counterterm of the form
$$
   {1\over 8\pi}\sum_i c_i\, M_i^2 \ln {M_i^2\over \mu^2} \int d^2x\,\sqrt{g}
$$
in the original action.  We could also avoid the need for the counterterm
by imposing the additional Pauli-Villars condition
\begin{align}
   m^2 \ln {m^2\over \mu^2} + \sum_i c_i\, M_i^2 \ln {M_i^2\over \mu^2} = 0.  
\label{PV3}
\end{align}
If we do this, note that 
to satisfy the three Pauli-Villars conditions (\ref{PV1}), (\ref{PV2})
and
(\ref{PV3}) with each $c_i = 1$ or $c_i = -2$, while keeping 
the ability to take the masses $M_i \to \infty$, we will need at least
five auxiliary fields.  If we use exactly five, three of these should be
bosonic and two Grassmann.    

We now consider the case of a curved space,  
where we need to calculate the limit
$$
  \expect{T^{\,i}_{\phantom{i}i}}_g = 2\pi\lim_{\Lambda\to\infty}\left\{m^2\, G_{m,\Lambda} (x,x)
    + \sum_i c_i\,M_i^2\, G_{M_i,\Lambda} (x,x)\right\} \expect{1}_g.
$$
Here $G_{m,\Lambda} (x,x)$ denotes the two-point function of a scalar field
computed with large-momentum cutoff $\Lambda$.  
The Pauli-Villars conditions are precisely what are needed to make 
the limit $\Lambda \to \infty$  finite. 

In a background $g_{ij}$, the propagator $G_{M} (x,x')$ satisfies
$$
  -\partial_i\left[\sqrt{g}\,g^{ij}\partial_j\, G_M(x, x')\right]
   + \sqrt{g}\,M^2\,G_M(x, x') = \delta (x-x').
$$
Given $x'$, we can change coordinates so that 
$x'$ lies at the origin.
In two dimensions, we can furthermore choose the coordinate system so that
\begin{align*}
  g_{ij} (x) &= e^{2\omega}\,\delta_{ij}, \\
  \omega (0) &= 0.
\end{align*}
so that the equation for the propagator becomes
\begin{align*}
  \left(-e^{-2\omega (x)}\Delta + M^2\right)\, G_M(x, 0) &= e^{-2\omega
  (x)}\delta(x) \\
&= e^{-2\omega (0)}\delta(x) \\
      &= \delta (x),
\end{align*}
since $\omega (0) = 0$.
Here $\Delta \equiv \delta^{ij}\partial_i\partial_j$.  As we increase the 
Pauli-Villars mass $M$, the propagator becomes increasingly short-ranged, 
and we will develop an expansion for $G_M(x)$ in terms of derivatives of 
$\omega$ at the origin.  
Expanding
around $x^i = 0$, we have 
\begin{align*}
  e^{-2\omega(x)} = 1 - 2\,(\partial_i \omega)\, x^i
    + \bigl[-(\partial_i \partial_j\omega) +
    2 \,(\partial_i \omega)\,(\partial_j \omega)\bigr]  x^i x^j  + \cdots,
\end{align*}
where the derivatives are all evaluated at $x^i = 0$.  We can then write
$$
  G_M = {1\over A - B} = {1\over A} +  {1\over A}\,B\, {1\over A}
     +  {1\over A}\,B\, {1\over A}\,B\, {1\over A} + \cdots,
$$
where $A \equiv -\Delta + M^2$, so that $1/A$ is the flat space propagator,
while 
$$
  B \equiv - \left \{2\,(\partial_i \omega)\, x^i
    + \bigl[(\partial_i \partial_j\omega) -
    2 \,(\partial_i \omega)\,(\partial_j \omega)\bigr] x^i x^j + \cdots
   \right\} 
    \Delta.
$$
In the first-order contribution $A^{-1} BA^{-1}$ to $G_M(0,0)$, 
the term with odd 
integrand proportional to $x^i$ vanishes.  The next term in  $A^{-1}
BA^{-1}$ is proportional to 
\begin{align*}
   \int d^2x &\,x^i x^j
     \int {d^2k\over (2\pi)^2}\, {e^{ikx}\over k^2 + M^2} 
     \int {d^2p\over (2\pi)^2}\, {p^2\,e^{-ipx}\over p^2 + M^2} \\
   &= -\int d^2x 
     \int {d^2k\over (2\pi)^2} \int {d^2p\over (2\pi)^2}\, 
       {1\over k^2 + M^2}\,{p^2\over p^2 + M^2}\,
      \partial_{k_i} \partial_{k_j} e^{i(k - p) x} 
      \\
 &= -(2\pi)^2 
     \int {d^2k\over (2\pi)^2} \int {d^2p\over (2\pi)^2}\, 
       {1\over k^2 + M^2}\,{p^2\over p^2 + M^2}\,
      \partial_{k_i} \partial_{k_j} \delta(k - p) 
      \\
&= -
     \int {d^2k\over (2\pi)^2}\,
    \left\{ 
      {-2 \delta^{ij}\over (k^2 + M^2)^2} + {8k_ik_j\over (k^2 + M^2)^3} 
    \right\} 
       {k^2\over k^2 + M^2}  \\
&= -
     \delta^{ij}\int {d^2k\over (2\pi)^2}\,
    \left\{ 
      {-2\over (k^2 + M^2)^2} + {4 k^2\over (k^2 + M^2)^3} 
    \right\} 
       {k^2\over k^2 + M^2}
      \\
&= 
     -2\,\delta^{ij}\int {d^2k\over (2\pi)^2}\,
      {k^2\,(k^2 - M^2)\over (k^2 + M^2)^4}
      \\
&=  -{1\over 12\pi}\, {1\over M^2}\, \delta^{ij}.
\end{align*}
The next nonzero term in  $A^{-1} BA^{-1}$
has integrand proportional to $x^ix^jx^kx^l$.  This may be checked to be
of order $1/M^4$.  Further terms are of even higher order in $1/M$, so that
the full first-order contribution is 
$$
  G_{M,1}(0, 0) = -{1\over 12\pi}\, {1\over M^2}\, 
     \left[\Delta\omega -
    2 \,\partial_i \omega\partial^i \omega\right] + o\left({1\over M^4}\right),
$$
which will contribute a term of zeroth order in 
$1/M^2$ to the expectation value $\expect{T^i_{\phantom{i} i}}_g$.

There is one additional contribution of order $1/M^2$ to $G_M(0,0)$.  It comes
from the second-order term $A^{-1}BA^{-1}BA^{-1}$ in the above expansion.
It is 
\begin{align*}
   &4\,(\partial_i \omega)\,(\partial_j \omega)
    \int d^2x \int d^2y \,x^i y^j \times \\
    &\quad \times
     \int {d^2k\over (2\pi)^2}\, {e^{ikx}\over k^2 + M^2} 
     \int {d^2p\over (2\pi)^2}\, {p^2\,e^{ip(y-x)}\over p^2 + M^2} 
     \int {d^2q\over (2\pi)^2}\, {q^2\,e^{-iqy}\over q^2 + M^2}. 
\end{align*}
By similar manipulations, this becomes
\begin{align*}
   -8M^2\,\delta^{ij}\,(\partial_i \omega)\,(\partial_j \omega)
     \int {d^2k\over (2\pi)^2}\, {k^4\over (k^2 + M^2)^5}
   &=  -{1\over 6\pi}\, {1\over M^2} \,\partial_i \omega\partial^i \omega, 
\end{align*}
so that 
\begin{align*}
  G_{M, 2} (0,0) = - {1\over 6\pi}\, {1\over M^2} \,\partial_i \omega\partial^i \omega
    + o\left({1\over M^4}\right)
\end{align*}
Notice that this contributes a term that exactly cancels the
term of the same form in $G^1_M(0,0)$.  
We find
\begin{align*}
  G_M^\Lambda (0,0) &=  {1\over 4\pi} \ln {\Lambda^2\over M^2}
       -
       {1\over 12\pi}\, {1\over M^2}\, 
                  \Delta\omega +  o\left({1\over M^4}\right) \\
  &=  {1\over 4\pi} \ln {\Lambda^2\over M^2}
       +
       {1\over 24\pi}\, {1\over M^2}\, 
                  R +  o\left({1\over M^4}\right).
\end{align*}
The matter contribution can be expanded in $m$ by writing the equation
for the propagator as 
\begin{align*}
  \left(-\Delta + e^{2\omega (x)}\,m^2\right)\, G_m(x, 0) &= \delta(x), 
\end{align*}
so that
$$
  G_m = {1\over A - B} = {1\over A} +  {1\over A}\,B\, {1\over A}
     +  {1\over A}\,B\, {1\over A}\,B\, {1\over A} + \cdots
$$
where $A \equiv -\Delta + m^2$ and
$$
  B \equiv m^2\left(1 - e^{2\omega (x)}\right) \equiv m^2 \gamma (x).
$$
Therefore
\begin{align*}
  G_m^\Lambda (0, 0) &=  {1\over 4\pi} \ln {\Lambda^2\over m^2} \\
                &\quad+ m^2 \int d^2x\, {1\over 2\pi}\,K_0(m|x|)\,\gamma(x) \,{1\over 2\pi}\,K_0(m|x|) \\
               &\quad+ m^4 \int d^2x\int d^2y\, {1\over 2\pi} \,K_0(m|x|)\,\gamma(x)\,
                {1\over 2\pi}\, K_0(m|y-x|) \,\gamma(y)\, {1\over 2\pi}\,
                K_0(m|y|) \\
     &\quad+ \cdots,
\end{align*}
where the Bessel function 
$$
  {1\over 2\pi}\, K_0(m|x|) = \int {d^2k\over (2\pi)^2}\, {e^{ikx}\over k^2 + m^2}
$$
decays exponentially for large $x$ and
$$
  K_0(m|x|) 
  \to - \ln (m|x|)
$$
as $m|x|\to 0$, so that the integrals are well-defined for a large
class of functions
$\gamma(x)$.  Since 
$$
\lim_{m\to 0} m \ln (m|x|) \to 0,
$$
all terms but the first in the above series vanish as $m\to 0$ as long
as $\gamma(x)$ is sufficiently well-behaved 
that we may exchange the integral and
the limit.  We find 
$$
    G_m^\Lambda (0, 0) \to {1\over 4\pi} \ln {\Lambda^2\over m^2}
$$
as $m\to 0$.  Including the Pauli-Villars contributions, we find 
\begin{align*}
  \expect{T^i_{\phantom{i}i}}_g &=  \half\, \left(m^2 \ln {\Lambda^2\over m^2}
         + 
       \sum_i c_i\, M_i^2 \ln {\Lambda^2 \over M_i^2}\right)
   - {1\over 12}\, 
                  R + o\left({1\over M^2}\right),
\end{align*}
where we have used the Pauli-Villars condition (\ref{PV1}) to obtain
the negative sign for the curvature term.  
As before, the terms in parentheses can be compensated by a counterterm or
made zero by the condition (\ref{PV3}) on the Pauli-Villars masses.  
After doing this, we 
find, as we set out to prove, that
\begin{align}
  \expect{T^i_{\phantom{i}i}}_g &\to
   -{1\over 12}\, 
                  R 
\end{align}
as the masses $m \to 0$ and $M_i \to \infty$. 
As a result, (\ref{PVvar}) becomes   
\begin{align}
  \delta \int [d\phi]_g^{PV}\,e^{-S(g, \phi, \bar\chi_i, \chi_i)}
    &= \left({1\over 24\pi}\int d^2x\,\sqrt{g}\,(\delta\omega)\,R
    \right)\expect{1}_g.
\label {PVvariation}
\end{align}
This agrees with the formula (\ref{confanom}) calculated in the 
Fujikawa approach, but here the anomaly comes from the Pauli-Villars
mass terms, not the measure, which is invariant under $\delta$.

\section{A study of $T_{zz}$}

It is instructive to observe the effect of the 
Pauli-Villars regularization on the expectation values 
of $T_{zz}$ and $T_{\bar z\bar z}$.  
As a side benefit, we shall identify the nontrivial relationship
between the energy-momentum tensor $T_{ij}$ 
in the path integral formalism, which is a true coordinate
invariant and finite tensor object, and the anomalous nontensor 
object in the operator formalism.  
By identifying explicitly the path integral expression 
corresponding to the latter, we shall obtain
an explanation, within the context of the path integral, 
for the nontensor transformation property.

Writing $z\equiv x^1 + i x^2$, we have
$$
  \expect{T_{zz}}_g = -2\pi\lim_{\Lambda\to\infty}\left\{ 
           \partial_z \partial_w G_{m}^{\Lambda} (z,w)
    + \sum_i \partial_z\partial_w G_{M_i}^{\Lambda}
    (z,w)\right\}_{w\to z}.
$$
In general, the matter contribution $\expect{T_{zz}^{m=0}}$ may diverge in the cutoff
parameter, and this divergence will already be compensated by the zeroth-order
contribution (the term $1/A$ of the previous section) in the 
expansion around flat space in conformal coordinates 
$$
\sum_i \expect{T_{zz}^{M_i}}_0 \equiv
-2\pi\sum_i c_i\,\partial_z\partial_w \left<z|A^{-1}|w\right>_{w\to z}
$$
 of the Pauli-Villars fields.  We
shall see explicit examples of this the next section.
So although the full $\expect {T_{zz}}_g$ is \textit{finite} and 
\textit{covariant}, 
let us also introduce a name for the combination
\begin{align}
  \hat T_{zz}\equiv 
    T_{zz}^{m=0} + \lim_{M_i\to\infty} \sum_i
    \expect{T_{zz}^{M_i}}_0.
\label{renorm}
\end{align}    
This may be regarded as defining a particular minimal
subtraction renormalization prescription.  But note that it is 
\textit{coordinate-dependent}.  
As we shall see, it is this combination that will turn out to correspond to the
renormalized energy-momentum operator occurring in the operator
formalism.  Obviously, it differs from the full energy-momentum tensor
$T_{zz}$ of our system.  But it only differs by a \textit{finite} amount.    

The Pauli-Villars contributions may be expanded in $1/M^2$ as in our calculation
of the trace anomaly in the previous section.  
 We find first-order contributions of the form 
\begin{align*}
   &-2\pi\,\bigl[-(\partial_i \partial_j\omega) +
    2 \,(\partial_i \omega)\,(\partial_j \omega)\bigr] \times \\
    &\quad \times \int d^2x\,x^i x^j 
    \, {1\over 2\pi}\,\partial_z K_0(M|x|)\, {1\over 2\pi}\,\partial_z \Delta K_0(M|x|).
\end{align*}
The term with integrand proportional to $\bar z \bar z$ vanishes
\begin{align*}
  \int d^2x &\,\bar z \bar z 
    \, \partial_{z} K_0(Mr)\, \partial_{z} \Delta K_0(Mr) \\
   &= M^2 \int d^2x \,\bar z \bar z \bar z \bar z 
    \, \partial_r K_0(Mr)\, \partial_r \Delta K_0(Mr) \\
   & = 0,
\end{align*}
since the integrand is odd under $z \to e^{i\pi/4}\, z$.  
Similarly, the term with integrand proportional to $z\bar z$ vanishes,
and we are left with the term  
\begin{align*}
  \int d^2x &\,z^2 
    \,{1\over 2\pi}\, \partial_{z} K_0(Mr)\, {1\over 2\pi}\,\partial_{z} \Delta K_0(Mr) \\
   &= \int d^2x \,z^2
     \int {d^2k\over (2\pi)^2}\, {\half\,\bar k \,e^{\half \,i\,(k\bar z+ \bar k z)}\over k\bar k + M^2} 
     \int {d^2p\over (2\pi)^2}\, {\half\,\bar p \,(p\bar p)\,e^{-\half \,i\,(p\bar z+ \bar p z)}\over p\bar p + M^2} \\
 &= - (2\pi)^2
     \int {d^2k\over (2\pi)^2} \int {d^2p\over (2\pi)^2}\, 
       {\bar k\over k\bar k + M^2}\,{\bar p\,(p\bar p)\over p \bar p + M^2}\,
      \partial_{\bar p}^2\, \delta^2(k - p) \\
 &= 
     -\int {d^2k\over (2\pi)^2} \, 
       {\bar k\over k\bar k + M^2}\,\partial_{\bar k}^2
     \left({\bar k\,(k\bar k)\over k \bar k + M^2}\right),
\end{align*}
which is of the form $\int d^2k\, f \left(\partial_{\bar k}^2 g\right)$.
In the last step, we could just as well have used the identity 
$$\partial_{\bar p}^2\, \delta^2(k - p) = \partial_{\bar k}^2\, 
\delta^2(k - p)$$
to write this in the form $\int d^2k\, \left(\partial_{\bar k}^2 f\right)g$.
However, the reader may check by explicit calculation that 
$$
  \int d^2k\, f \left(\partial_{\bar k}^2 g\right)
   \ne \int d^2k\, \left(\partial_{\bar k}^2 f\right)g,
$$
an ambiguity due to naively exchanging the integral over $x$ 
with those over $k$ and $p$.  In general, integrals may be exchanged only
after a careful analysis of their uniformity of convergence.\cite{whittaker}  
Here we 
see an order dependence proportional to the quantities
$
  f \left(\partial_{\bar k} g\right)
$
or
$
  \left(\partial_{\bar k} f\right) g
$
integrated over the surface of the integration region.  Since these
surface integrands are of order $1/r$, they do contribute in two dimensions
(the corresponding surface terms may be checked to be of order
$1/r^2$ in our previous calculation of the trace anomaly, and therefore
did not threaten the validity of that calculation).  Fortunately we 
can avoid these uniformity 
issues by using the equation
of motion $(-\Delta + M^2)\,{1\over 2\pi}\, K_0(Mr) = \delta^2(x)$ 
in position space first as follows: 
\begin{align*}
  \int d^2x &\,z^2 
    \, {1\over 2\pi}\,\partial_{z} K_0(Mr)\, {1\over 2\pi}\,\partial_{z} \Delta K_0(Mr) \\
  &=\int d^2x\,z^2 
    \, {1\over 2\pi}\,\partial_{z} K_0(Mr)\, \partial_{z} \biggl(-\delta^2(x) + {1\over 2\pi}\,M^2\,K_0(Mr)\biggr) \\
&= M^2\int d^2x\,z^2 
    \, {1\over 2\pi}\,\partial_{z} K_0(Mr)\, {1\over 2\pi}\,\partial_{z} K_0(Mr) 
\end{align*}
where we have used the behaviour $\partial_z K_0(Mr)= o(1/r)$ as $r\to 0$ to 
drop the term containing $\partial_{z} \delta^2(x)$.  This then becomes
\begin{align*}
   M^2 \int d^2x &\,z^2
     \int {d^2k\over (2\pi)^2}\, {\half\,\bar k \,e^{\half \,i\,(k\bar z+ \bar k z)}\over k\bar k + M^2} 
     \int {d^2p\over (2\pi)^2}\, {\half\,\bar p\,e^{-\half \,i\,(p\bar z+ \bar p z)}\over p\bar p + M^2} \\
 &= -M^2
     \int {d^2k\over (2\pi)^2} \, 
       {\bar k\over k\bar k + M^2}\,\partial_{\bar k}^2
     \left({\bar k\over k \bar k + M^2}\right) \\
 &= 2M^2
     \int {d^2k\over (2\pi)^2} \left\{
       {k\bar k\over \left(k\bar k + M^2\right)^2}\ - 
     {(k\bar k)^2\over \left(k \bar k + M^2\right)^4} \right\} \\
  &= {1\over 12\pi}.
\end{align*}
The first-order contribution is therefore
\begin{align*}
   \expect{T_{zz}^M}_1 &= -{1\over 6}\,\left[-(\partial_z^2\omega) +
    2 \,(\partial_z \omega)^2\right] 
   + o\left({1\over M^2}\right).
\end{align*}
The only other contribution of order $0$ in $1/M^2$ is the second-order
term
\begin{align*}
  -2\pi\cdot 4&\left(\partial_z\omega\right)^2
   \int d^2x \int d^2 y\, z w\,{1\over 2\pi}\,\partial_z K_0(M|x|)\, 
       {1\over 2\pi}\,\Delta K_0(M|y - x|)\, {1\over 2\pi}\,\partial_w \Delta K_0 (M|y|) \\
  &= 
       - 8\pi\left(\partial_z\omega\right)^2
   \int d^2x \int d^2 y\, z w\,{1\over 2\pi}\,\partial_z K_0(M|x|)\, 
       \bigl(-\delta(y-x) + {1\over 2\pi}\,M^2 K_0(M|y - x|)\bigr)\times \\
  &\qquad\qquad\qquad\qquad\qquad\qquad\times\partial_w
       \bigl(-\delta(y) +{1\over 2\pi}\, M^2 K_0(M|y|)\bigr) \\
  &= - 8\pi\left(\partial_z\omega\right)^2 \biggl\{
   M^4\int d^2x \int d^2 y\, z w\,{1\over 2\pi}\,\partial_z K_0(M|x|)\, 
       {1\over 2\pi}\,K_0(M|y - x|)\,{1\over 2\pi}\,\partial_w K_0(M|y|) \\ 
   &\qquad\qquad\qquad - M^2
   \int d^2x\, z^2\,{1\over 2\pi}\,\partial_z K_0(M|x|)\, 
       {1\over 2\pi}\,\partial_z K_0(M|x|)\biggr\},
\end{align*}
where again we have manipulated the integrals in position space
into a form where the surface terms in momentum space will vanish.  
This becomes
\begin{align*}
  -8\pi&\left(\partial_z\omega\right)^2
   \biggl\{
   -M^4 \int {d^2 k\over (2\pi)^2}\,
            \partial_{\bar k}\left({\bar k\over k\bar k+ M^2}\right)
            \,{1\over k\bar k+ M^2}\,
            \partial_{\bar k}\left({\bar k\over k\bar k+ M^2}\right) \\
   &\qquad\qquad
    + M^2\int {d^2 k\over (2\pi)^2}\,
            \left[\partial_{\bar k}\left({\bar k\over k\bar k+ M^2}\right)
       \right]^2
   \biggr\} \\
  &\qquad = {1\over 6}\left(\partial_z\omega\right)^2,
\end{align*}
so that the second order contribution is
\begin{align*}
   \expect{T_{zz}^M}_2 &= {1\over 6}\,(\partial_z \omega)^2
   + o\left({1\over M^2}\right).
\end{align*}
As $M\to \infty$, we get for each Pauli-Villars field
\begin{align*}
   \expect{T_{zz}^{M_i}} &= \expect{T_{zz}^{M_i}}_0 + {1\over 6}\,c_i\,(\partial_z \omega)^2. 
\end{align*}
Including the matter contribution and using the Pauli-Villars
condition
(\ref{PV}), we find
\begin{align}
   \expect{T_{zz}} &= \expect{\hat T_{zz}} 
     - {1\over 6}\,\left[(\partial_z^2\omega) -
    \,(\partial_z \omega)^2\right] \\
   &\equiv \expect{\hat T_{zz}} - {1\over 12}\, t_{zz}.
\label {thatt}
\end{align}
Since 
$
  \Gamma^z_{zz} = 2\,\partial_z\omega
$,
we may write $t_{zz}$ in the form
$$
  t_{zz} = 
  \partial_z\Gamma^z_{zz}-\half\left(\Gamma^z_{zz}\right)^2
$$
used by Eguchi and Ooguri in the context of the axiomatic 
approach to conformal field theory.\cite{eguchi,ooguri}  
One may verify that 
$t_{zz}$ transforms as
\begin{align}
  \delta_v  t_{zz} \equiv   \,\partial_z^3 v^z
                      + v^i \partial_i  t_{zz} 
  + 2 \,(\partial_z v^z)\, t_{zz}
\end{align}
under the action of a holomorphic vector field $v$.   
Since $T_{zz}$ is a component of a true
tensor, the term
$\langle\hat T_{zz}\rangle$ in (\ref{thatt}) is therefore not a tensor in isolation.  From the transformation laws for $T_{zz}$ and $t_{zz}$, 
it follows that 
 $\langle\hat T_{zz}\rangle$ 
transforms as
\begin{align}
  \delta_v  \expect{\hat T_{zz}} =   {1\over 12} \,\partial_z^3 v^z
                      + v^i \partial_i \expect{\hat T_{zz}} + 2 
    \,(\partial_z v^z)\,\expect{\hat  T_{zz}},
\end{align}
which coincides with the transformation law 
obtained in the operator product formalism
after point-splitting 
renormalization.\cite{eguchi,ooguri,BPZ,polchinski,gawedzki,henkel,alvarez} 
The above result provides a straightforward explanation,
in the context of
the path integral formalism, for the familiar
anomalous transformation law of $\hat T_{zz}$.  In the path integral
formalism, the more natural object is in fact the true tensor $T_{zz}$, which
includes the full Pauli-Villars correction and, as we shall see,
satisfies simpler Ward identities.  The non-covariant split (\ref{thatt}) is 
rather less natural from this point of view.

Note that it would be incorrect to try to absorb the Pauli-Villars
effects into an operator redefinition 
by replacing
$$
  T_{zz} \to \hat T_{zz} - {1\over 12}\,t_{zz}
$$
in general calculations of expectation values.  For example, with such a redefinition
one would lose essential contact terms in the Ward identities that 
we consider in section \ref{wardsection}.

The above transformation law for $t_{zz}$ can be integrated to give, for  $z' = f(z)$,
$$
  t_{z'z'} \,dz'\otimes dz' = t_{zz}\, dz\otimes dz -
      \{f, z\}\, dz\otimes dz,
$$ 
where 
$$
  \{f, z\} \equiv -6 \lim_{w\to z} \left(
       {f'(w)\,f'(z)\over \left(f(w) - f(z)\right)^2}
   - {1 \over (w - z)^2}
     \right),
$$
is called the Schwarzian derivative.  In this form, the transformation
is easily seen to be precisely the difference between the renormalization
subtractions needed in the operator product formalism for energy-momentum
tensors defined in different coordinate systems, which provides a simple
way to see that $\hat T_{zz}$ corresponds to the usual operator 
formalism definitions.

Defining $t_{z\bar z}= t_{\bar z z} = 0$ and
$
  t_{\bar z \bar z} = 
  \partial_{\bar z}\Gamma^{\bar z}_{\bar z\bar z}
   -\half\left(\Gamma^{\bar z}_{\bar z \bar z}\right)^2
$,
we may summarize the results of this section as 
\begin{align}
\expect{T_{ij}}_g = \langle\hat T_{ij}\rangle_g 
                    - {1\over 12}\,t_{ij} - {1\over 24}\,g_{ij}R.
\label{PIfromOP} 
\end{align}

\section{Examples}

In this section we provide some example computations of  
energy-momentum tensor expectation values on a couple of simple manifolds. 

First we consider the plane, for which
$$
 \expect {T_{zz}^m}_g =  
     - {2\pi\over 4} \int {d^2 k\over (2\pi)^2} \,{(k_1 - ik_2)^2\over k^2 + m^2}.
$$    
The numerator in the integrand is $k_1^2 - 2 i k_1 k_2 - k_2^2$.  The cross
term is odd in $k_1$ and $k_2$ separately, while the sum of the 
first and the last term
is odd under $k_1 \leftrightarrow k_2$, so that 
the integral vanishes 
when integrated over the particular choice  
$0 \le k\le \Lambda$ of cutoff region in momentum space, suitable
 for manifolds for which $k_1$ and 
$k_2$ are continuous.  With this cutoff region, we get the contribution
$$
  \expect {T_{zz}^{m=0}} = 0.
$$
Similarly,
$$
   \expect {T_{zz}^{M_i}} = 0.
$$
for each Pauli-Villars field, and we find
$$
  \expect {T_{zz}} = 0 =   \expect {\hat T_{zz}}.
$$
In section \ref{PV} we showed, using a Pauli-Villars definition of
the path integral measure, that the expectation value
$$
  \expect{T_{z \bar z}} = -{1\over 48}\,R = 0
$$
on the plane.

In a later section, we will also need the expectation value 
$$
  \expect{\partial_z \phi\, \partial_{\bar z}\phi}
    =  {1\over 4}\int {d^2 k\over (2\pi)^2} \,{k^2\over k^2 + m^2}.  
$$
This integrand is not odd, so that our simple argument for $T_{zz}$ 
cannot be used.  Computing this with a large-momentum cutoff $\Lambda$,
we find
$$
   \expect{\partial_z \phi\, \partial_{\bar z}\phi}
    = {1\over 16\pi} \, \Lambda^2 - {1\over 16\pi} \,m^2\,\ln {\Lambda^2
       \over m^2}.
$$
As we saw, both these terms can be exactly canceled by 
introducing Pauli-Villars fields with the appropriate statistics and
mass conditions, and we obtain 
$$
   \expect{\partial_z \phi\, \partial_{\bar z}\phi + \mathit{PV}}  
     = 0
$$
on the plane.

Let us employ the Pauli-Villars formalism to perform a path-integral 
calculation of the expectation value $\expect{T_{zz}}$ on an infinite
cylinder of circumference $L$, 
a simple nontrivial manifold with a length scale that
will give rise to a Casimir energy.  We have, for a field of mass $M$,
\begin{align*}
   \expect{T_{zz}^M} &= - 2\pi \cdot {1\over 4}\cdot{1\over L}\cdot\sum_n
                \int {dk_2\over 2\pi}\, 
                {\left({2\pi n\over L}\right)^2 - k_2^2 \over 
                  \left({2\pi n\over L}\right)^2 + k_2^2 + M^2}    
                 \\
    &= -{\pi\over 2}\cdot {1\over L}\sum_n\left\{
                 \sqrt {\left(\textstyle{2\pi n\over L}\right)^2 + M^2}
                      - {M^2/2\over 
                  \sqrt {\left({2\pi n\over L}\right)^2 + M^2}}
                 \right\}.
\end{align*}
We have performed the integral over $k_2$ by closing the integration
contour either above or below the real line.  Although the resulting 
contour integral strictly diverges as we move the contour to infinity,
this is one term in   
the full integrand containing both the matter and all Pauli-Villars,
the integral of which 
will converge due to the Pauli-Villars conditions, validating
the contour integral argument.  

We first evaluate the matter contribution, for which $M = 0$.  Inserting 
a convergence factor, the dependence on 
which will be canceled by the Pauli-Villars
contributions, we find 
\begin{align*}
   \expect{T_{zz}^{m=0}} 
    &= -{\pi\over L}\sum_{n>0}
                 {2\pi n\over L} \, e^{-\epsilon\,{2\pi n/ L}}\\
   &= -{\pi\over L}\cdot{2\pi\over L}\,
        {e^{-\epsilon\, 2\pi/L}\over \left(1 - e^{-\epsilon\, 2\pi/L}\right)^2}.
\end{align*}
Using the expansion
$$
  {e^{-\nu}\over \left(1 - e^{-\nu}\right)^2} = {1\over\nu^2}
      - {1\over 12} + o\left(\nu\right),
$$
we get
\begin{align*}
   \expect{T_{zz}^{m=0}} 
    &= -{1\over 2\,\epsilon^2} + {1\over 24}\,\left({2\pi\over L}\right)^2 +
       o\left(\epsilon\right).
\end{align*}
It remains to show that, as promised, the Pauli-Villars fields will 
cancel the dependence on $\epsilon$ as $\epsilon\to 0$.  As $M\to\infty$,
we can write each Pauli-Villars contribution as a continuous integral
as follows. 
\begin{align*}
   \expect{T_{zz}^M} 
    &= -{1\over 2}\cdot M^2\cdot {2\pi\over ML}\sum_{n\ge 0}\left\{
                 \sqrt {\left(\textstyle{2\pi n\over ML}\right)^2 + 1}
                      - {1/2\over 
                  \sqrt {\left({2\pi n\over ML}\right)^2 + 1}}
                 \right\} e^{-\epsilon M\,{2\pi n/ ML}} \\
    &\to -{1\over 2}\cdot M^2 \int_0^\infty dx\,\left\{
                 \sqrt {x^2 + 1}
                      - {1/2\over 
                  \sqrt {x^2 + 1}}
                 \right\} e^{-\epsilon M\,x} \\
   &= -{1\over 2}\cdot M^2\cdot {\sqrt{\pi}\over 2}
                 \,\biggl\{
                   {2\over \epsilon M}\,\Gamma \left({3\over 2}\right)
                    \left[\mathbf{H}_1(\epsilon M) - N_1(\epsilon M)\right]\\
   &\qquad\qquad\qquad\qquad\quad
                    + \half\,\Gamma \left({1\over 2}\right)
                    \left[\mathbf{H}_0(\epsilon M) - N_0(\epsilon M)\right]
                   \biggr\},
\end{align*}
where $\mathbf{H}_\nu$ denotes Struwe functions and $N_\nu$ Bessel functions
of the second kind.\cite{GR}  Using the small $z$ expansions
\begin{align*}
  \mathbf{H}_0 (z) &= o \left(z\right), \\
  \mathbf{H}_1 (z) &= o \left(z^2\right), \\
  \pi \,N_0 (z) &= 2 \left(\ln{z\over 2} + \mathbf{C}\right) 
                + o \left(z^2\right) \\
  \pi N_1 (z) &= -{2\over z} - {z\over 2} + z \left(\ln{z\over 2} + \mathbf{C}\right)
                + o \left(z^3\right), 
\end{align*}
we find
\begin{align*}
   \expect{T_{zz}^M} 
    &= - {1\over 2 \epsilon^2} - {1\over 8} \, M^2 + o(\epsilon).
\end{align*}
By the Pauli-Villars conditions $\sum c_i = 1$ and 
$\sum c_i M_i^2 = 0$ from (\ref{PV1}) and (\ref{PV2}), 
the first term cancels the $\epsilon$ dependence 
of the matter field and the second term falls away. 
Our final answer for the full expectation value becomes
$$
  \expect {T_{zz}} = {1\over 24}\,\left({2\pi\over L}\right)^2.
$$
This is finite without further
renormalization.

\section{Reparametrization-invariance of the measure}

Consider a family of reparametrizations 
$f_\lambda : M \to M$, $f_0 = \mathrm{id}$
where $\lambda$
is a real parameter, acting
on scalar fields via push-forward
$$
  \phi(z) \to \phi^\lambda (z) \equiv \phi (f_{-\lambda}(z)).
$$ 
In this section, we will study the transformation of the path 
integral measure under such deformations acting only on the fields,
\textit{not on the metric}.  
Given a metric $g$ on $M$, let $\phi_n$ be a basis of field configurations
 satisfying
$$
  \delta_{mn} = \int d^2x \,\sqrt{g} \,\phi_m(x) \,\phi_n(x)
$$
and expand $\phi^\lambda$ as
$$
  \phi^\lambda (x) \equiv \sum_n a^\lambda_n\, \phi_n(z),  
$$
For each $\lambda$, the map $\phi \mapsto (a^\lambda_0, a^\lambda_1, \ldots)$
may be regarded as a coordinate chart on the space of field configurations $\phi$.
Denoting the $\infty$-form
$$
  \bigwedge_n da^\lambda_n \equiv da^\lambda_0 \wedge da^\lambda_1 \wedge \cdots
  \equiv [d\phi^\lambda],
$$
the change of variables theorem tells us that
$$
  \int [d\phi^\lambda] \, e^{-S(\phi^\lambda)}
    = \int [d\phi] \, e^{-S(\phi)}.
$$
Note that his had better be true if the concept of integration is to make
sense in a chart-independent way.
In preparation for our derivation of the Ward identities in the next 
section, let us determine how the form $[d\phi^\lambda]$
depends on $\lambda$.  
We can project out the coefficient
$$
  a^\lambda_m = \int d^2x\,\sqrt{g} \,\phi_m(x)\,
                       \phi^\lambda(x),
$$
and calculate
\begin{align*}
  {d\over d\lambda}\, a^\lambda_m
   &= \int d^2x\, \sqrt{g}\,
                    \phi_m(x)\,
                       (- v^i)\partial_i\phi^\lambda(x) \\
   &= -\sum_n a^\lambda_n \int d^2x \,\sqrt{g}\,\phi_m (x)\,
              v^i\partial_i\phi_n(x)
      \\
   &\equiv \sum_n C_{mn}\, a^\lambda_n
\end{align*}
where $v^i$ is the vector field whose flow gives $f_\lambda$, in other words
$$
   v^i(x)\equiv \left.{df^{\,i}_\lambda(x)\over d\lambda}\right|_{\lambda = 0}
$$  
The normal rules for manipulating differential forms now give
$$
  {d\over d\lambda}\, [d\phi^\lambda] \equiv
   {d\over d\lambda}\,\bigwedge_n da^\lambda_n 
   = \left(\sum_m C_{mm}\right) \,\bigwedge_n da^\lambda_n =  \left(\Tr\, C\right) \,[
     d\phi^\lambda].
$$
Now
\begin{align*}
 C_{mm} &= -
 \int d^2x \,
       \sqrt{g}\,\phi_m(x)\, v^i\partial_i \,\phi_m(x) \\
  &= \int d^2x \,\sqrt{g}\left(v^i\partial_i \phi_m (x)\right)
       \phi_m(x) \\
  &\phantom{=} + \int d^2x \,\phi_m (x)\,\partial_i\left(
       \sqrt{g}\, v^i\right)\phi_m(x) \\
  &= \half\int d^2x \,\phi_m (x)\,\partial_i\left(
       \sqrt{g}\, v^i\right)\phi_m(x)  \\
 &= \half\int d^2x \,\sqrt{g}\,\phi_m (x)\left(\nabla_i
        v^i\right)\phi_m(x) 
\end{align*}
where $\nabla_i$ denotes the covariant derivative on the worldsheet with 
respect to the deformed metric $f^*_\lambda g$.  
Similar to the calculation of the Weyl anomaly, we get the trace of an
operator $\nabla_i\, v^i$ representing 
a change in area, in this case along the flow of $v$.  The calculation of 
the trace can be done via a heat kernel regularization and is
identical to the calculation of the Weyl anomaly, with 
$\nabla_i v^i$ replacing $2 \, \delta\omega$.
We get
\begin{equation*}
  {d\over d\lambda}\, [d\phi^\lambda]
  = \left( 
      {1 \over 8\pi\epsilon} \int d^2x \, \sqrt{g}\,
                    \nabla_i v^i
  + {1\over 48\pi} \int d^2x \, \sqrt{g}
          \left( \nabla_i v^i\right) R_{g}
    \right)  [d\phi^\lambda]
\end{equation*}
The integrand $ \sqrt{g}\, \nabla_i v^i = 
    \partial_i\left(\sqrt{g}\,v^i\right)$ in the first term 
is a total derivative, and will integrate to zero for suitable boundary 
conditions on $v^i$, which we will assume.  
A similar partial integration in the second term, and 
remembering that $\nabla_i = 
\partial_i$ on scalars, leads to
\begin{equation}
  {d\over d\lambda}\, \left[d\phi^\lambda\right]
  = \left(-{1\over 48\pi} \int d^2x \, \sqrt{g}\,
           v^i\,\nabla_i R_{g}
     \,\right) [d\phi^\lambda].
\label{anom}
\end{equation}
In particular, on a flat manifold, the 
measure will be invariant under reparametrizations of $\phi$.  It is also 
interesting to note that the prefactor is independent of 
the deformation parameter $\lambda$.   

As in the section 3, a similar analysis applies to the 
Pauli-Villars auxiliary fields.  As a result, the full measure
is reparametrization-invariant:
\begin{align}
  {d\over d\lambda}\, \left[d\phi^\lambda\right]_{\mathit{PV}}
  &= \left(1 + \sum_i c_i\right) \left(-{1\over 48\pi} \int d^2x \, \sqrt{g}\,
           v^i\,\nabla_i R_{g}
     \,\right) [d\phi^\lambda]_{\mathit{PV}} \nonumber \\
  &= 0.
\label{reparaminvariance}
\end{align}

\section{Ward identities}
\label{conservesection}

In this section we derive the Ward identity for covariant conservation
of energy-momentum.  At first we do not use Pauli-Villars fields, and
we therefore need to keep careful account of the transformation of the 
measure in the spirit of Fujikawa, obtaining an anomalous conservation law.  
We then discuss why this procedure is not quite correct, motivating
the introduction of a Pauli-Villars measure.  The Pauli-Villars measure
transforms trivially, but now the energy-momentum tensor has extra terms
making the insertion finite.  We find that 
the full, finite energy-momentum insertion satisfies the 
classical conservation
law, which we then show to be consistent with the anomalous 
conservation law for the corresponding
quantity in the operator formalism.  

The starting point for deriving the energy-momentum conservation Ward identities  
is the change of variables theorem\cite{polchinski}
\begin{equation}
  \int [d\phi^\lambda] \, e^{-S(\phi^\lambda)}
    = \int [d\phi] \, e^{-S(\phi)},
\label{change1}
\end{equation}
where
\begin{align*}
  S(\phi^\lambda) & = \half\int d^2x \, \sqrt g\, g^{ij} \partial_i \phi^\lambda\partial_j\phi^\lambda, \qquad \phi^\lambda(z) = \phi (f_{-\lambda} (z)).
\end{align*} 
Note that the deformation $\phi \to \phi^\lambda$ does not act on 
the metric. 

The above formula extrapolates a well-known property of finite-dimensional
integration to the infinite-dimensional case.  It says that the 
path integral is independent of the choice of coordinates on 
the space of fields, an essential property that must be satisfied by any 
reasonable definition of integration.  In chapter 9 we will 
perform a consistency check on this formula for our particular choice.

Classically, covariant conservation of energy-momentum is a consequence 
of the observation that, since the action density in $S$ is 
a coordinate-invariant expression, it would be invariant if the  
deformation $f_\lambda$ acted not only on the fields but also on
the metric.  However, since the deformation
acts only on the fields, the variation of $S$ is not in general zero but 
instead proportional to the omitted variation
$\delta g^{ij}\equiv\nabla^i v^j + \nabla^j v^i$ 
of the metric
(see the calculation of $dS/d\lambda$ below).  
This allows one to obtain a Noether current via standard methods,
even though the corresponding conservation law involves covariant
derivatives and cannot in general be integrated to give
conserved charges.  

We now provide a path integral derivation of the current by  
varying (\ref{change1}) 
with respect to a generic deformation $f_\lambda$.  
Differentiating with respect to $\lambda$, the right hand side gives zero.
For the left hand side, we have already calculated the variation of 
$[d\phi^\lambda]$.  The variation of $S$ is
\begin{align*}
  {dS\over d\lambda} 
  &= \half\cdot{d\over d\lambda}
      \int d^2x \, \sqrt g\, g^{ij} \partial_i \phi^\lambda\partial_j\phi^\lambda \\
   &=  \half\cdot{d\over d\lambda}\int d^2x \,
       \sqrt {f^*_\lambda g}\, (f^*_\lambda g)^{ij}  
        \partial_i \phi\,\partial_j\phi &&\textit{(by pullback)}\\
  &= -{1\over 4\pi} \int d^2x\,\sqrt {f^*_\lambda g}\, 
              {d\over d\lambda}(f^*_\lambda g)^{ij} \,T^{f^*_\lambda g}_{ij}(\phi) 
        &&\textit{(definition  of $T$)}\\
   &=  {1\over 4\pi} \int d^2x\,\sqrt {f^*_\lambda g}\, 
              (\nabla^i_{f^*_\lambda g} v^j  +\nabla^j_{f^*_\lambda g} v^i) 
         \,T^{f^*_\lambda g}_{ij}(\phi) && (\mathcal L_v g^{ij} = 
  - \nabla_g^i v^j - \nabla_g^j v^i) \\
   &=  {1\over 4\pi} \int d^2x\,\sqrt {g}\, 
              (\nabla^i v^j + \nabla^j v^i) 
         \,T^g_{ij}(\phi^\lambda) &&\textit{(by push-forward)}\\
  &= {1\over 2\pi}\int d^2x\,\sqrt {g}\, 
              (\nabla^i v^j) \,T_{ij}(\phi^\lambda)&&\textit{(symmetry of $T$)} \\
  &= {1\over 2\pi}\int d^2x\,\sqrt {g}\, \left(
              \nabla^i (v^j \,T_{ij}(\phi^\lambda)) 
             -  v^j \,\nabla^i T_{ij}(\phi^\lambda)\right) \\
  &=  -{1\over 2\pi}\int d^2x\,\sqrt {g}\, v^j \,\nabla^i T_{ij}(\phi^\lambda),
\end{align*}
where 
$$
  T^{g}_{ij}(\phi^\lambda) \equiv -2\pi\left(\partial_i\phi^\lambda\partial_j\phi^\lambda
             - \half\,
          g_{ij}g^{kl}\,\partial_k\phi^\lambda\,\partial_l\phi^\lambda\right).
$$
and where we have used the property 
$\sqrt g\,\nabla^i J_i = \partial^i (\sqrt g\, J_i)$ of the covariant derivative 
to get a
total derivative which we have dropped in the last line.  
Differentiation of equation (\ref{change1}) with respect to $\lambda$ 
then gives
\begin{align*}
  0 &=
     \int d^2x\,  \sqrt{g} 
      \expect{{1\over 2\pi}\,v^j\,\nabla^i\,T_{ij}(\phi^\lambda) - {1\over 48\pi} \,
           v^i\,\nabla_i R}_\lambda \\
  &= {1\over 2\pi}\int d^2x\,  \sqrt{g} \, v^j\, \nabla^i
      \expect{T_{ij}(\phi^\lambda) - {1\over 24} \,
           g_{ij} R}_\lambda
\end{align*}
Since $v^j$ is arbitrary, we get the conservation law
\begin{equation}
  \nabla^i
      \expect{T_{ij}(\phi^\lambda) - {1\over 24} \,
           g_{ij} R}_\lambda = 0,
\label{conserve}
\end{equation}
where the variation of the measure has contributed a curvature 
term not present in the classical conservation law.  
Remembering our definition (\ref{tildeT}) of $\tilde T_{ij}$, this 
is the same as
$$
   \nabla^i\expect{ \tilde T_{ij}(\phi^\lambda)}_\lambda = 0.
$$
In this derivation, which assumed a single massless scalar field for
which  $T^i_{\phantom{i}i}$ is identically zero,  
the curvature term
in the conservation law was due to the variation of the path integral 
measure under a reparametrization.  

As it stands, though, this formula
is not quite meaningful.  In fact, as we have seen,
the expectation values $\expect{T_{ij}}$ for a single matter field
are not generally finite, and 
we needed to introduce Pauli-Villars fields to obtain
finite values for the full energy-momentum, including matter and 
auxiliary fields, 
in a coordinate-invariant way.  
As shown in (\ref{reparaminvariance}),
the combined matter-Pauli-Villars path integral 
measure is invariant under 
reparametrizations, and 
we find by the above argument that
\begin{equation}
\nabla^i
      \expect{T_{ij}^\lambda}_\lambda \equiv
  \nabla^i
      \expect{T_{ij}(\phi^\lambda) + \mathit{PV}}_\lambda = 0.
\label{conservePV}
\end{equation}
In other words, the full $T_{ij}^\lambda$ satisfies the 
classical covariant conservation law.  Note that, by (\ref{reparaminvariance}),
we may now drop the subscript $\lambda$ denoting the measure used to
calculate the expectation values.  Repeating  the definition 
(\ref{PIfromOP}) of $\hat T_{ij}$,
$$
\expect{T_{ij}^\lambda} \equiv \expect{\hat T_{ij}^\lambda} - {1\over 12}\, t_{ij} - {1\over 24}\,g_{ij}R,
$$  
we
see that the quantity $\hat T_{ij}$, which corresponds to the 
operator formalism energy-momentum, does satisfy an
anomalous conservation law depending on the curvature, consistent
with the corresponding operator-formalism calculations.

How about conformal symmetry?  A conformal vector field satisfies
$\delta g^{ij}\equiv-\nabla^i v^j - \nabla^j v^i - 2\,\delta\omega\, g^{ij} = 0$. 
By symmetry of $T_{ij}$ and 
the product rule for the covariant derivative, we find from 
 (\ref{conservePV}) and
 (\ref{PVanom}) that
\begin{equation}
  \nabla^i
      \expect{v^j T_{ij}^\lambda} = - \,\delta\omega\,
    \expect{T^{\,i}_{\phantom{i}i}} = {1\over 12}\,\delta\omega\, R.
\end{equation}
If the right hand side were zero, $v^j T_{ij}^\lambda$ would span
 an infinite-dimensional
family of covariantly conserved currents.  Instead, these currents
have an effective source proportional to the local conformal factor and 
the curvature.  
In other words, background curvature 
breaks conformal invariance.  Since curvature gives an intrinsic local scale
to a manifold that can be observed by quantum diffusion processes, 
this is not surprising.

\section{The Ward identity for $T_{ij}$, or, where is the anomaly?}
\label{wardsection}

In this section we study the Ward identities for the transformation
law of the energy-momentum tensor.   
We confirm by explicit calculation that the full energy-momentum tensor
satisfies the \textit{classical\/} Ward identities, and 
we explain how these are related to the anomalous Ward identities found
in the operator formalism.  

In the process we perform a very careful, direct path-integral 
calculation of nontrivial contact terms arising
in expectation values of certain energy-momentum products, 
previously derived only using axiomatic considerations.  
The results of this section will then be used in the following section
to  perform a consistency check supporting the validity of the change of
variables formula for the infinite-dimensional
integration used to define the path integral.  
The contact terms will also be essential to
our review of the 
relationship between the conformal anomaly and the energy-momentum 
two-point functions in the last section.

The Ward identity for an insertion of $T_{ij}$ follows from the
change of variables
\begin{equation}
  \int [d\phi^\lambda]_\mathit{PV} \,T_{ij}^\lambda(x)\, 
   e^{-S(g, \phi^\lambda, \bar\chi_i^\lambda, \chi_i^\lambda)}
    = \int [d\phi]_{\mathit{PV}} \,T_{ij}(x)\, e^{-S(g, \phi, \bar\chi_i, \chi_i)}.
\label{change}
\end{equation}
Here $T_{ij}^\lambda$ includes the contribution of the Pauli-Villars
auxiliary fields.    Since $[d\phi^\lambda]_\mathit{PV}$ is independent of $\lambda$, as 
explained in section \ref{conservesection}, we find the following
Ward identity after
differentiating both sides with respect to $\lambda$,
\begin{align}
  \expect{{d\over d\lambda}\, T_{ij}^\lambda(x)}
   + {1\over 4\pi}\int d^2y\, \sqrt{g} \, h^{kl}(y)\expect{T_{ij}^\lambda (x) T_{kl}^\lambda(y)} = 0,
\label{Ward}
\end{align}
where
$$
  h^{kl} \equiv -\nabla^k v^l - \nabla^l v^k.
$$
and
\begin{align}
  {d\over d\lambda}\, T_{ij} 
    = -\mathcal{L}_v T_{ij} 
      &+ \pi\, \mathcal{L}_v \left(g_{ij}g^{kl}\right)
         \left(\partial_k\phi\,\partial_l\phi 
             + \sum_i \partial_k\bar\chi\,\partial_l\chi) \right) \nonumber\\
      &+ \pi\, \mathcal{L}_v g_{ij}
        \left(m^2\phi^2 + \sum_i M_i^2 \bar\chi_i\chi_i \right),
\label{lie}
\end{align}
where $\mathcal{L}_v$ denotes the Lie derivative.  For example,
$$    
   \mathcal{L}_v T_{ij}  = v^i\partial_i T_{ij} + (\partial_i v^m)\, T_{mj}
   + (\partial_j v^m)\, T_{im},
$$
and where $\mathcal{L}_v (g_{ij}g^{kl}) \equiv 
\mathcal{L}_v (g\otimes g^{-1})_{ij}^{\phantom{ij}kl}$.
The second and the third term in (\ref{lie}) subtract the contributions to 
$\mathcal{L}_v T_{ij}$ coming from varying the metric, since in
$d T_{ij}^\lambda/d\lambda$ only the fields are varied, not the
metric. 

In the special case where the vector field $v$ is holomorphic in 
a neighbourhood of the
point $x$ of the insertion (though not necessarily everywhere), generating
a conformal deformation of
 that neighbourhood, the 
second term in (\ref{lie}) vanishes.  In this case, and only
in this case, we may write,
in a local conformal coordinate system with $g_{ij} = e^{2\omega}\delta_{ij}$,
\begin{align}
  {d\over d\lambda}\, T_{zz} 
   & = -v^i\partial_i T_{zz} - 2 (\partial_z v^z)\, T_{zz} \label{Tzzconformal}\\
  {d\over d\lambda}\, T_{z\bar z} 
   &=  -v^i\partial_i T_{z\bar z} + v^i \partial_i (2\omega)\, T_{z\bar z}. 
   \label{Tzbarzconformal}
\end{align}

It is important to note that the second term in the Ward identity (\ref{Ward})
generates the transformation $\mathcal{L}_v T_{ij}$ for 
the components $T_{zz}$ and $T_{\bar z \bar z}$ only when 
$v$ is holomorphic in a neighbourhood of the insertion.  In other cases, the
extra terms in (\ref{lie}) cannot be ignored.

   It is also important to note that the transformation $\mathcal{L}_v T_{ij}$
appearing in the Ward identity is the 
\textit{classical} one.  This 
 expresses the fact that, since the Pauli-Villars regularization is coordinate-invariant, the full energy-momentum tensor $T_{ij}$,
including Pauli-Villars contributions, is finite and a
true coordinate invariant tensor quantity.  As discussed in section 4, 
it is the quantity $\hat T_{zz}$ introduced in (\ref{renorm}) 
that satisfies an anomalous transformation law.  We shall see by explicit
calculation that the above, non-anomalous 
Ward identity is indeed correct, but contains contact terms
that can be compensated by a redefinition of $T_{ij}$ to obtain the 
familiar anomalous Ward identity for the modified insertion $\hat T_{zz}$.

For simplicity, we restrict attention to the plane and consider
\begin{align*}
    \expect{{d\over d\lambda}\, T_{zz}^\lambda}
  =
    &- {1\over 4\pi}\int d^2w \, h^{w w}\expect{T_{zz}^\lambda\,
        T_{ww}^\lambda} \\
    &- 2 \cdot {1\over 4\pi} \int d^2w \, h^{w \bar w}\expect{T_{zz}^\lambda\,
        T_{w\bar w}^\lambda} \\
    &- {1\over 4\pi}\int d^2w \, h^{\bar w \bar w}\expect{T_{zz}^\lambda\,
        T_{\bar w\bar w}^\lambda} 
\end{align*}
The first expectation value on the right hand side is easily calculated 
by a double contraction to be
$$
\expect{T_{zz}^\lambda\,
        T_{ww}^\lambda}
   = \half\,{1 \over (z - w)^4},
$$
where self-contractions vanish due to the Pauli-Villars conditions 
as in section 5.  
The expectation values in the second and third terms above are not discussed
in many standard treatments, but in fact contribute contact 
terms, 
in the absence of which the above identity would be untrue.
The presence of the contact terms are inferred using axiomatic frameworks
in Refs.~\refcite{gawedzki}, \refcite{cappelli1} and
\refcite{cappelli2}, 
but we have been unable
to find a calculation from first principles as presented below.  

The contact terms are nontrivial to calculate.
Consider for example 
$\expect{T_{zz}\, T_{\bar w\bar w}}$.  Naively taking  
appropriate derivatives of the double contraction for a massless field
would give the square of the delta function, which does not 
exist as a well-defined distribution.    
Also troublesome is
the expectation value $\expect{T_{zz}\, T_{w\bar w}}$.  
Since $T_{w\bar w}$ is identically zero in a massless theory, one 
might expect the answer to be zero.  With our careful 
definition of the path integral, we shall see that this is 
only true up to a contact term.  

Let us therefore calculate these expectation values more carefully
using our regularized path integral. 
We realize an infrared 
regularization by introducing a mass $m$ for the field $\phi$, eventually
to be taken to zero, while the ultraviolet regularization is taken
care of, as before, by the Pauli-Villars 
auxiliary fields whose masses we eventually take to infinity.

We start by considering the expectation value 
$\expect{T_{zz}\, T_{\bar w\bar w}}$. 
Writing the contractions in terms of derivatives of the propagator
$$
  \expect{\phi(x)\,\phi(0)} = \int {d^2p\over (2\pi)^2}\,
     {e^{-ipx} \over p^2 + m^2}
$$
and Fourier transforming the result gives
the familiar one-loop Feynman integral
\begin{align*}
  \expect{T_{zz}(x)\, T_{\bar z\bar z}(0)} &=
    {2 \,(2\pi)^2\over 16}\int {d^2p\over(2\pi)^2}\, e^{-ip\cdot x} 
        \int {d^2k\over(2\pi)^2}\,\left\{F(m) + \sum_i c_i \,F(M_i)\right\} 
\end{align*}
where 
\begin{equation}
  F(m) \equiv {k^2 \, (p- k)^2\over \left[k^2 + m^2\right]\,
              \left[(p - k)^2 + m^2\right]}.
\end{equation}
In the absence of the Pauli-Villars field contributions, the  
integral over $k$ would have both a quadratic and a logarithmic divergence.  
The regularization consists in choosing the coefficients $c_i$ and 
masses $M_i$ so as to make the integral finite.  Assuming this has been
done, we can then write, using identities such as $k^2 = k^2 + m^2 - m^2$, 
 \begin{align*}
  \int {d^2k\over(2\pi)^2}\, &\left\{
        {k^2 \, (p- k)^2\over \left[k^2 + m^2\right]\,
              \left[(p - k)^2 + m^2\right]} + \mathit{PV}\right\}
        \\
   &=   \int {d^2k\over(2\pi)^2}\,
         \biggl\{1 -
        {m^2\over
              (p - k)^2 + m^2} - 
          {m^2\over
              k^2 + m^2}   \\
         &\qquad\qquad\qquad\quad +  {m^4\over \left[k^2 + m^2\right]\,
              \left[(p - k)^2 + m^2\right]}
         + \mathit{PV}\biggr\}  \\
   &=   \int {d^2k\over(2\pi)^2}\,
         \biggl\{1 -
        {2\,m^2\over
              k^2 + m^2} 
         +  {m^4\over \left[k^2 + m^2\right]\,
              \left[(p - k)^2 + m^2\right]}
         + \mathit{PV}\biggr\},
\end{align*}  
where the shift of $k$ in the last line is permitted since the 
integral converges.    
Integrating the first two terms 
between $0$ and $\Lambda$, we get,
for large $\Lambda$, 
\begin{align*}
  {1\over 4\pi}\,\Lambda^2 - {1\over 2\pi}\,m^2\ln {\Lambda^2\over m^2} 
         +  \int {d^2k\over(2\pi)^2}\, {m^4\over \left[k^2 + m^4\right]\,
              \left[(p - k)^2 + m^2\right]}
         + \mathit{PV}
\end{align*}  
By the Pauli-Villars conditions
(\ref{PV1}), (\ref{PV2}) and (\ref{PV3}),
\begin{align*}
  1+ \sum_i c_i &= 0,  \\
  m^2 + \sum_i c_i\,M_i^2 &= 0, \\
 m^2\ln {m^2\over \mu^2} + \sum_i c_i \,M_i^2 \ln {M_i^2\over \mu^2}
  &= 0,
\end{align*}
the first two terms cancel entirely.\cite{anselmi}
What remains is the finite integral
\begin{align*}
   \expect{T_{zz}(x)\, T_{\bar z\bar z}(0)} = {2 m^4\,(2\pi)^2\over 16}
    \int {d^2p\over(2\pi)^2}\, & e^{-ip\cdot x}
     \int {d^2k\over(2\pi)^2}\, { 1\over \left[k^2 + m^2\right]}\,
              {1\over\left[(p - k)^2 + m^2\right]} \\
         &+ \mathit{PV}
\end{align*}
Feynman's trick gives
\begin{align*}
   \int {d^2k\over(2\pi)^2}\,&{1\over k^2 + m^2}\,
               {1\over (p - k)^2 + m^2} \\
   &=  \int {d^2k\over(2\pi)^2}
     \int_0^1 dx\, {1\over \bigl\{x\,( k^2 + m^2) + (1-x)\,[(p - k)^2 + m^2]\bigr\}^2} \\
   &= \int {d^2k\over(2\pi)^2}
     \int_0^1 dx\,{1\over \left(k^2 + x(1-x)\,p^2 + m^2\right)^2},
\end{align*}
where we have redefined $k - (x - 1)\,p\to k$ to complete the 
square in the last line.  Performing the straightforward integration over 
$k$, we find
$$
   {1\over 4\pi}\int_0^1 dx\,{1\over x(1-x)\,p^2 + m^2},
$$   
Changing variables from $x$ to
$$
  s \equiv {m^2\over x \,(1 - x)},
$$
and including the prefactor, we find the spectral representation\cite{cappelli2}
\begin{align*}
  {1\over 4\pi}\cdot {2\, (2\pi)^2\over 16} &\int_{4m^2}^\infty {ds \over s}\,
    {  2\, m^4\over \sqrt{1 - {4m^2/ s}}}\cdot {1 \over p^2 + s} \\
 &= {1\over 16}\cdot 4\pi\int_{2m}^\infty {d\mu \over \mu}\,
    {2\,m^4\over \sqrt{1 - {4m^2/ \mu^2}}}\cdot {1 \over p^2 + \mu^2} \\
  &= {1\over 16}\cdot{\pi\over 3}
      \int_{2m}^\infty {d\mu\over \mu^5}\,
  {24\,m^4\over \sqrt{1 - {4m^2/ \mu^2}}}\cdot {1 \over p^2 + \mu^2} \\
  &= {1\over 16}\cdot{\pi\over 3}
      \int d\mu\, c(\mu, m) 
         \cdot {\mu^4  \over p^2 + \mu^2}
\end{align*} 
for the Fourier transform of the expectation value.  Here the spectral 
function
$$
   c(\mu, m) \equiv  {24\, m^4\over \mu^5\,\sqrt{1 - {4m^2/ \mu^2}}}\,
         \theta (\mu - 2m)
$$ 
is dimensionless and has area equal to $1$, independent
of $m$.  
Contributions to the expectation value come from
two-particle intermediate states propagating between $0$
and $x$.   The lowest of these has energy $2m$, which
explains the lower bound on the integral.

To confirm that the area is one, we calculate\cite{GR}
\begin{align*}
  24\, m^4 \int_{2m}^\infty {d\mu\over\mu^5}\,{1\over \sqrt{1 - 4m^2/\mu^2}}
    &= {3\over 2} \int_{1}^\infty 
         {d\eta\over\eta^4}\,{1\over \sqrt{\eta^2 - 1}} \\
   &= {3\over 2}\cdot\half\cdot B(2, \hhalf)\\ 
   & = 1.
\end{align*}
As 
$m\to 0$, $c(\mu, m)$ develops a spike at $2m$ and goes to zero elsewhere.
It
follows that
$$
  c(\mu, m) \to\delta(\mu) \quad \textrm{as} \quad m\to 0.
$$
The Fourier transformed expectation value, including the Pauli-Villars
contributions, is then 
\begin{align*}
{1\over 16}\cdot{\pi\over 3} &
      \int d\mu\, \left\{ c(\mu, m) + \sum_i c_i\, c(\mu, M_i) \right\} 
         \cdot {\mu^4  \over p^2 + \mu^2} \\
  &= {1\over 16}\cdot{\pi\over 3}
      \int d\mu\, \left\{ c(\mu, m) + \sum_i c_i\, c(\mu, M_i) \right\} 
         \cdot \left(\mu^2 - p^2 + {p^4 \over p^2 + \mu^2} \right)
\end{align*}
A change of variables from $\mu$ to $\eta$ as above shows that the
contribution 
$$\int d\mu\, \left\{ c(\mu, m) + \sum_i c_i\, c(\mu, M_i) \right\} \mu^2$$
is proportional to $m^2 + \sum_i c_i \, M_i^2$, which is zero by the 
Pauli-Villars conditions.  So is the contribution
$$\int d\mu\, \left\{ c(\mu, m) + \sum_i c_i\, c(\mu, M_i) \right\} p^2,$$ 
which is proportional to $1 + \sum_i c_i$ by the fact that $c(\mu,\cdot)$ 
has unit area.  We are therefore left with 
\begin{align*}
  {1\over 16}\cdot{\pi\over 3} 
      \int d\mu\, \left\{ c(\mu, m) + \sum_i c_i\, c(\mu, M_i) \right\} 
         \cdot  {p^4 \over p^2 + \mu^2}.
\end{align*}
Here, as we take the Pauli-Villars masses to infinity, 
we find
$$
\int_{2M_i} d\mu\, c(\mu, M_i) 
         \cdot  {p^4 \over p^2 + \mu^2} \to 0\quad \textrm{as}\quad M_i\to\infty
$$
because of the lower bound on the integration and the unit area property of
$c(\mu, \cdot)$ making the integrand of order
$1/M_i^2$.  All that remains is the matter contribution which, as we remove
the infrared cutoff, is 
$$
{1\over 16}\cdot{\pi\over 3}
      \int d\mu\, c(\mu, m)  
         \cdot {p^4 \over p^2 + \mu^2} \to {1\over 16}\cdot{\pi\over 3} 
    \cdot p^2 \quad \textrm{as} \quad m\to 0,
$$ 
since in this limit $c(\mu, m) \to \delta (\mu)$.  Fourier transforming, 
we find
\begin{align}
\expect{T_{zz}(x)\, T_{\bar z\bar z}(0)} &\to
  - {\pi\over 12}\, \partial_z \partial_{\bar z}\delta(x)
\label{zzbarzbarz}
\end{align}
as $m\to 0$.

It is important to point out that, in addition to the above contribution,
we would expect additional terms due to self-contractions.  However, as discussed in
section 5, these all vanish on the plane.

Next we calculate the expectation 
value
$$
  \expect{T_{z\bar z}(z)\, T_{z\bar z}(0)}.
$$
Since the mass term breaks conformal invariance, $T_{z\bar z}$ is not zero.
In fact
$$
  T_{z\bar z} = {\pi\over 2}\,m^2 \,\phi^2.
$$
Although this indeed goes identically to zero as $m\to 0$, 
a contact term survives in the limit $m\to 0$.  
Indeed, 
\begin{align*}
  \expect{T_{z\bar z}(x)\, T_{z\bar z}(0)} =
    2m^4 \,\left(\pi\over 2\right)^2\int {d^2p\over(2\pi)^2}\, e^{-ip\cdot x} 
        \int {d^2k\over(2\pi)^2}\,\biggl\{&{1\over k^2 + m^2}\,
               {1\over (p - k)^2 + m^2} \\
   &+ \textit{PV}\biggr\}, 
\end{align*}
which is the same expression we obtained above for
$\expect{T_{zz}(x)\, T_{\bar z\bar z}(0)}$.  Therefore
\begin{align}
  \expect{T_{z\bar z}(x)\, T_{z\bar z}(0)} 
    &\to
  - {\pi\over 12}\, \partial_z \partial_{\bar z}\delta(x)
\label {zbarzzbarz}
\end{align}
as $m\to 0$.

The remaining expectation value $\expect{T_{zz}(x)\, T_{z\bar z}(0)}$
has Fourier transform (with a slight abuse of notation we denote
$\bar k \equiv k_1 - i k_2$
but keep $k^2 = k_1^2 + k_2^2$) 
\begin{align*}
 -{1\over 4}&\cdot 2m^2 \cdot (-2\pi)\cdot\left(\pi\over 2\right)\int {d^2k\over(2\pi)^2}\, \left\{
        {\bar k \, (\bar p- \bar k)\over \left[k^2 + m^2\right]\,
              \left[(p - k)^2 + m^2\right]} + \mathit{PV}\right\} \\
   &=  {2m^2\,(2\pi)^2\over 16}\int {d^2k\over(2\pi)^2}\int_0^1 dx\left\{
          {\left[\bar k + (1 - x) \bar p\right]\,
           \left[\bar p - \bar k - (1 - x) \bar p\right]
          \over [k^2 + x(1-x)\,p^2 + \mu^2]^2} + \mathit{PV}\right\} \\
    &=  {2m^2\,(2\pi)^2\over 16}\int {d^2k\over(2\pi)^2}\int_0^1 dx\left\{
          {x (1 - x) {\bar p}^2
          \over [k^2 + x(1-x)\,p^2 + \mu^2]^2} + \mathit{PV}\right\}, 
\end{align*}  
where in the last line we have dropped odd integrands, including terms 
proportional to $k_1^2 - k_2^2$ and $k_1k_2$.  Performing the 
integration over $k$ and changing variables from $x$ 
to $\mu$ as before, we find
\begin{align*}
   {1\over 16}\cdot{\pi\over 3}
      & \int_{2m}^\infty {d\mu\over \mu^5}\,
  {24\,m^4\over \sqrt{1 - {4m^2/ \mu^2}}}\cdot {\mu^2\,{\bar p}^2
       \over p^2 + \mu^2}  + \textit{PV}\\
  &= {1\over 16}\cdot{\pi\over 3}
      \int d\mu\, \left\{c(\mu, m) 
         \cdot {\mu^2 \,{\bar p}^2  \over p^2 + \mu^2}
        + \textit{PV}\right\} \\
   &= {1\over 16}\cdot{\pi\over 3}
      \int d\mu\, \left\{c(\mu, m)\cdot {\bar p}^2 - c(\mu, m)\cdot
          {p^2 \,{\bar p}^2  \over p^2 + \mu^2}
        + \textit{PV}\right\}.
\end{align*} 
As before, the first integrand will cancel due to the Pauli-Villars 
condition $1 + \sum_i c_i$ = 0,
while the second term will vanish for the Pauli-Villars fields in the 
limit of infinite mass.  Again, all that remains is the matter 
contribution
\begin{align*}
   - {1\over 16}\cdot{\pi\over 3}
      \int d\mu\,  c(\mu, m)\cdot
          {p^2 \,{\bar p}^2  \over p^2 + \mu^2} \to 
    - {1\over 16}\cdot{\pi\over 3}\, {\bar p}^2
\end{align*} 
as $m\to 0$.  Fourier transforming, we get
\begin{align}
\expect{T_{zz}(x)\, T_{z\bar z}(0)}
   &\to
   {\pi\over 12}\, \partial_{z}^2\delta(x)
\label{zzzbarz}
\end{align}

Summarizing, we have 
\begin{align}
\expect{T_{zz}(x)\,
        T_{zz}(0)}
   &= \half\,{1 \over (z - w)^4}, \\
\expect{T_{zz}(x)\, T_{\bar z\bar z}(0)} &=
  - {\pi\over 12}\, \partial_z \partial_{\bar z}\delta(x), \\
 \expect{T_{z\bar z}(x)\, T_{z\bar z}(0)} 
    &=
  - {\pi\over 12}\, \partial_z \partial_{\bar z}\delta(x), \\
  \expect{T_{zz}(x)\, T_{z\bar z}(0)}
   &=
   {\pi\over 12}\, \partial_{z}^2\delta(x).
\end{align}

The same formulae were obtained in the axiomatic approach to conformal field
theory in Ref.~\refcite{gawedzki}.  A separate argument provided in  
Ref.~\refcite{cappelli2} motivates them as follows: 
If we knew that conservation of
energy-momentum held even in the limit of coinciding points (which 
we actually do not know without explicit calculation), 
we could have expected the form of these correlation
functions by inserting a spectral 
decomposition of the unit operator between the two $T$s.  By
conservation of $T_{ij}$, the correlator must then have the form 
$$
  \expect{T_{\mu\nu}(x) T_{\rho\sigma}(0)}
   = {\pi\over 3} \int d\mu\,c(\mu)\int {d^2p\over (2\pi)^2}\,
   e^{ipx} {\left(g_{\mu\nu}p^2 - p_\mu p_\nu\right)
            \left(g_{\rho\sigma}p^2 - p_\rho p_\sigma\right)
       \over p^2 + \mu^2},
$$
which, noting that the Fourier transform of
$1/z^4$ is $(\pi/ 24)\, {\bar p}^4 / |p|^2$, indeed coincides with 
our results upon guessing $c(\mu) \propto \delta(\mu)$,
as expected for a massless theory.  Our calculation confirms
this explicitly.   Further work on contact terms of energy-momentum 
tensors is reported in 
Ref.~\refcite{forte}.

As a consistency check, even without performing the full calculation above, 
a simple algebraic argument applied
to the original integrands, combined with a
shift of variables, which is 
permitted in the presence of the Pauli-Villars fields,
shows that, for example,
$$
 \partial_{\bar z}\expect{ T_{zz} (x) T_{z \bar z}(0)} + 
   \partial_z\expect{ T_{\bar z z} (x) T_{z \bar z}(0)} = 0.
$$
In other words, the Pauli-Villars regularization
does not break conservation of energy-momentum, even in the limit of coinciding
points.

We are now finally ready to verify
 \begin{align*}
    \expect{{d\over d\lambda}\, T_{zz}^\lambda}
  =
    &- {1\over 4\pi}\int d^2w \, h^{w w}\expect{T_{zz}^\lambda\,
        T_{ww}^\lambda} \\
    &- 2 \cdot {1\over 4\pi} \int d^2w \, h^{w \bar w}\expect{T_{zz}^\lambda\,
        T_{w\bar w}^\lambda} \\
    &- {1\over 4\pi}\int d^2w \, h^{\bar w \bar w}\expect{T_{zz}^\lambda\,
        T_{\bar w\bar w}^\lambda} 
\end{align*}
Expressing the components of $h$ in terms of $v$, for example, 
$h^{ww} = -2(\partial_{\bar w}v^w + \partial_{\bar w}v^w)$,
discarding total derivative terms, and remembering that
$\partial_{\bar w} (1/w) = \pi\, \delta (w)$, we find
 \begin{align}
    \expect{{d\over d\lambda}\, T_{zz}^\lambda}
  &=
    - {1\over 12} \,\partial_z^3 v^z \nonumber\\
    &\quad\,+  {1\over 12} \,\partial_z^3 v^z + 
                 {1\over 12} \,\partial_{\bar z}\partial_z^2 v^{\bar z} \nonumber\\
    &\quad\,- {1\over 12} \,\partial_{\bar z}\partial_z^2 v^{\bar z} \nonumber\\
   &= 0
\label{wardcontact}
\end{align}
In other words, the contact terms neatly cancel the anomalous contribution 
coming from the $1/(z-w)^4$ term in 
 $\expect{T_{zz}^\lambda\, T_{ww}^\lambda}$.  
We may also calculate the left hand side directly.  We have,
by (\ref{lie}),
\begin{align*}
\expect{{d\over d\lambda}\, T_{zz}} & = 
-\mathcal{L}_v \expect {T_{zz}} 
      + \pi\, \mathcal{L}_v \left(g_{zz}g^{kl}\right)
         \expect{\partial_k\phi\,\partial_l\phi + \mathit{PV}}
      +  \pi\, \mathcal{L}_v g_{zz}
        \expect{m^2\phi^2 + \mathit{PV}} \\
& = 0,
\end{align*}
since all the expectation values on the right hand side
 were shown to vanish on the plane in sections 3 and 5.  
This confirms the validity of the classical Ward identity for the 
component
$T_{zz}$.  In the special case where $v$ is holomorphic in a 
neighbourhood of the insertion $T_{zz}$, this may be simplified using
(\ref{Tzzconformal}) and we find that we have verified the
expression
 \begin{align*}
    \expect{-v^i\partial_i T_{zz}^\lambda -2 \,(\partial_z v^z)\, T_{zz}^\lambda }
  =
    &- {1\over 4\pi}\int d^2w \, h^{w w}\expect{T_{zz}^\lambda\,
        T_{ww}^\lambda} \\
    &- 2 \cdot {1\over 4\pi} \int d^2w \, h^{w \bar w}\expect{T_{zz}^\lambda\,
        T_{w\bar w}^\lambda} \\
    &- {1\over 4\pi}\int d^2w \, h^{\bar w \bar w}\expect{T_{zz}^\lambda\,
        T_{\bar w\bar w}^\lambda}. 
\end{align*}
As promised, this Ward 
identity has \textit{no anomaly}.  

Let us also verify the Ward identity for an insertion of $T_{z\bar z}$.
A similar calculation to the above gives
 \begin{align*}
    \expect{{d\over d\lambda}\, T_{z\bar z}^\lambda}
  &=
    + {1\over 12} \,\partial_z^2 \partial_{\bar z} v^z \nonumber\\
    &\quad\,-  {1\over 12} \,\partial_z^2 \partial_{\bar z} v^z - 
                 {1\over 12} \,\partial_z \partial_{\bar z}^2 v^{\bar z} \nonumber\\
    &\quad\,+ {1\over 12} \,\partial_z \partial_{\bar z}^2 v^{\bar z} \nonumber\\
   &= 0
\end{align*} 
and, similar to the case of $T_{zz}$ above, the left hand side may also
be shown to be $0$ by the results of sections 3 and 5.

How does one reconcile our non-anomalous Ward identity for 
$T_{zz}$ with the anomalous identity appearing in the 
operator formalism literature?  
We note that if we define the quantity $\hat T_{zz}$ to 
coincide with $T_{zz}$
\begin{align}
  \hat T_{zz} = T_{zz} 
\end{align}
on the plane with trivial metric, and deform $\hat T_{zz}$ according
 to the 
 nontensor transformation law
\begin{align}
  \delta_v \hat T_{zz} \equiv   {1\over 12} \,\partial_z^3 v^z
                      + v^i \partial_i \hat T_{zz} + 2 \,(\partial_z v^z)\,\hat T_{zz},
\end{align}
as we deform the metric along the flow of a vector field $v$ holomorphic
in a neighbourhood of the insertion, the extra term in the 
transformation law of $\hat T_{zz}$ will exactly cancel the contributions
coming from the contact terms (second and third lines) in the derivation
(\ref{wardcontact}).
In terms of $\hat T_{zz}$, the Ward identity can therefore be expressed
as 
\begin{align*}
   - \expect{\delta_v \hat T_{zz}^\lambda}
     &=
    - {1\over 4\pi}\int d^2w \, h^{w w}\expect{\hat T_{zz}^\lambda\,
        \hat T_{ww}^\lambda} \\
    &=  {1\over\pi}\int d^2w \, (\partial_{\bar w} v^w)\expect{\hat T_{zz}^\lambda\,
        \hat T_{ww}^\lambda},
\end{align*}
which is the form familiar from the operator formalism.
Indeed, the energy-momentum tensor obtained from 
the common point-splitting renormalization of the operator 
product coincides with $\hat T_{zz}$, as can be seen from its 
transformation law. 

It is possible to express the metric-dependence of $\hat T_{zz}$ 
generated by the above transformation law directly 
as\cite{eguchi,ooguri}
\begin{align*}
  \hat T_{zz} = T_{zz} + {1\over 12}\,t_{zz},
\end{align*}  
where
$$
  t_{zz} \equiv \partial_z \Gamma^{z}_{\phantom{z}zz} 
      - \half\,\left(\Gamma^{z}_{\phantom{z}zz}\right)^2.
$$
Looking back to section 4, we see that $\hat T_{zz}$ here
coincides
with the corresponding $\hat T_{zz}$ of equation (\ref{thatt}).

Using the property\cite{eguchi}
$$
  \partial_{\bar z} t_{zz} = - \half\,g_{z\bar z}\,\partial_z R,
$$
the insertion  $\hat T_{zz}$ is easily seen to still satisfy the 
conservation law
$$
  \expect{\partial_{\bar z} \hat T_{zz}} = 0
$$
on a flat manifold.

\section{Checking the change of variables theorem}

Fundamental to path-integral derivations of conservation laws
and Ward identities is the generalization (\ref{change}) 
\begin{equation}
  \int [d\phi^\lambda]_\mathit{PV} \, e^{-S(g, \phi^\lambda, \bar\chi_i^\lambda, \chi_i^\lambda)}
    = \int [d\phi]_\mathit{PV} \, e^{-S(g, \phi, \bar\chi_i, \chi_i)}
\label{change*}
\end{equation}
of the change of 
variables theorem from finite to infinite dimensions.  
It says that the path integral is independent of the choice of coordinates 
on the space of fields, a statement that
must be satisfied if the concept of integration is to
make sense in a chart-independent way.  In the absence of 
a general proof, we here perform a small consistency check 
in support of this statement.   

In particular, let us
check this formula to second order around $\lambda = 0$ on the plane.
As explained in section \ref{conservesection}, $[d\phi^\lambda]_\mathit{PV}$ is
independent of $\lambda$,
so that differentiating the left hand side of (\ref{change*}) 
twice with respect to $\lambda$ gives
\begin{align*}
  \left.{d^2\over d\lambda^2}\right|_{\lambda = 0}
  \int [d\phi^\lambda]_\mathit{PV} \, e^{-S(g, \phi^\lambda, \bar\chi_i^\lambda, \chi_i^\lambda)} 
    &= {1\over 4\pi}\int d^2x\,h^{ij}(x)\,\expect{\left.{d\over d\lambda}\right|_{\lambda=0}\, T_{ij}^\lambda(x)}  \\
&\quad+  \left({1\over 4\pi}\right)^2 \int d^2x \int d^2y\, h^{ij}(x)\, h^{kl}(y) \expect{T_{ij}(x)\, T_{kl} (y)},
\end{align*}
where 
$$
  h^{ij} \equiv \left.\delta_\lambda(f^*_\lambda g)^{ij}\right|_{\lambda = 0}
 = 
    -\partial^i v^j - \partial^j v^i.
$$  
For the change of variables formula to be valid to second order, 
this expression should be
zero.  But notice that this expression is just the integral over 
$x$ of the Ward identity (\ref{Ward}), laboriously verified 
in the previous section.   
We therefore find the required result
\begin{align*}
  \left.{d^2\over d\lambda^2}\right|_{\lambda = 0}
  \int [d\phi^\lambda]_\mathit{PV} \, e^{-S(g, \phi^\lambda, \bar\chi_i^\lambda, \chi_i^\lambda)}
 &= 0.
\end{align*}

\section{Relating the conformal anomaly and the Ward identity}

By changing from an active to a passive point of view, 
the result of the previous section can also be interpreted as 
telling us that the second-order variation of the partition function is zero 
when we
pull the metric, as opposed to the fields, along the flow of a vector field.
Such a deformation does not change the curvature of an initially 
flat surface, and therefore, as expected, the Weyl anomaly did not contribute.  

Let us now instead consider the change of 
$$
  \int [d\phi]_g^\mathit{PV} \, e^{-S(g, \phi, \bar\chi_i, \chi_i)}
$$
to second order under a deformation
$h_{ij}\equiv\delta g_{ij}$ of the trivial metric $g_{ij} = \delta_{ij}$ 
on the plane, not necessarily generated by a vector field.  Since in general 
this cannot be compensated by
a change of variables, we do not expect the variation to be zero.  
In general, a second derivative will bring down up to two instances
of the energy-momentum tensor from the exponent, so that we will have 
to calculate terms of the form $\expect{T_{ij} T_{kl}}$, for which 
we are forced to use the Pauli-Villars regularization of the previous 
sections to obtain the correct contact terms.  These contact terms
are essential
to obtaining the correct result.

 Since the Pauli-Villars measure 
$[d\phi]_g^\mathit{PV}$ is invariant under variations of $g$, we can write
\begin{align*}
  \delta_g^2
  \int [d\phi]_g^\mathit{PV} \, e^{-S} 
   &=  {1\over 4\pi} \cdot \half \int d^2x\, \sqrt{g} \,g_{kl}\, h^{kl}\, h^{ij}
       \expect{T_{ij}} \\
    &\qquad + {1\over 4\pi}\int d^2x\,\sqrt{g} \,h^{ij}\expect{\delta_g T_{ij}}  \\
&\qquad +  \left({1\over 4\pi}\right)^2 \int d^2x \sqrt{g} \int d^2y\,\sqrt{g}\, h^{ij}(x)\, h^{kl}(y) 
    \expect{T_{ij}(x)\, T_{kl} (y)}.
\end{align*}
On the plane, $\sqrt g = 1$ and, as shown in section 5,
$\expect{T_{zz}}$, 
$\expect{T_{\bar z \bar z}}$, and $\expect{T_{z\bar z}}$ are all zero
with the Pauli-Villars measure,
so that the first term vanishes.
Remembering that $T_{ij}$ depends on the metric, we calculate 
\begin{align*}
\delta_g\expect{T_{ij}}
 &= 2\pi\cdot \half\,\delta_g(g_{ij}\,g^{kl}) \expect{\partial_k \phi\,
         \partial_l\phi + \mathit{PV}} + \pi \,\delta g_{ij}\, \expect{m^2\phi^2 + \mathit{PV}}.
\end{align*}
The expectation values on the right hand side 
are all linear combinations of $\expect{T_{zz}}$, 
$\expect{T_{\bar z \bar z}}$, $\expect{T_{z\bar z}}$
and $\expect{\partial_z\phi\,\partial_{\bar z}\phi + \mathit{PV}}$, all of which 
vanish on the plane as shown in sections 3 and 5.  So
\begin{align*}
\delta_g \expect{T_{ij}}
 &= 0.
\end{align*}
Let us now see how this Ward identity is related to the conformal 
anomaly discussed in sections 2 and 3.  
For simplicity we first consider a Weyl variation 
$$ 
  \delta_\omega g_{ij} = 2 \,\delta\omega \,g_{ij} 
$$
of the flat metric.  Then $h^{z\bar z} = h^{\bar z z} = -4\,\delta\omega$, and 
the above formula becomes
\begin{align*}
  \delta_\omega^2
  \int [d\phi]_g^\mathit{PV} \, e^{-S} 
    &= \left({1\over 4\pi}\right)^2 \int d^2z \int d^2w\, 4\,h^{z\bar z}\, 
           h^{w\bar w}\expect{T_{z\bar z} T_{w\bar w}}  \\
    &=  \left({1\over 4\pi}\right)^2 \int d^2z \int d^2w\, 4\,h^{z\bar z}\, 
           h^{w\bar w}\,\left(-{\pi\over 12}\right)\,
          \partial_z\partial_{\bar z} \delta(z - w) \, \expect{1} \\
    &= -{1\over 12\pi} \int d^2z \,\delta\omega\, \Delta\,\delta\omega
     \, \expect {1} \\
 &= {1\over 24\pi} \int d^2x \,\delta\omega\,\delta R\,\expect{1}, 
\end{align*}
where we used 
\begin{align*}
R &= -2\,e^{-2\omega}\Delta\omega \quad
    \textrm{for}\quad g_{ij} = e^{2\omega} \delta_{ij}.
\end{align*} 
This formula coincides precisely with the second variation around the flat 
metric ($R = 0$) of the 
formula (\ref{PVvariation}) for the conformal anomaly
$$
  \delta_\omega\int [d\phi]_g^\mathit{PV}\, e^{-S(g, \phi, \bar\chi_i, \chi_i)} = \left( {1\over 24\pi} \int
  d^2x\,\sqrt{g}\,\delta\omega(x)\,R \right) \int [d\phi]_g^\mathit{PV}\, e^{-S(g, \phi, \bar\chi_i, \chi_i)}.
$$
derived in section 3.  

For more generic deformations of the metric, we have
\begin{align*}
  \delta_g^2
  \int [d\phi]_g^\mathit{PV}  \, e^{-S} 
    &=  \left({1\over 4\pi}\right)^2 \int d^2x \sqrt{g} \int d^2y\,\sqrt{g}\, h^{ij}(x)\, h^{kl}(y) 
    \expect{T_{ij}(x)\, T_{kl} (y)}
 \\ 
&=  \left({1\over 4\pi}\right)^2 
   \int d^2z \int d^2w\,\biggl\{ h^{zz}\, h^{ww} \left(\half\right){1\over (z - w)^2} \\
  &\qquad\qquad\qquad\qquad\qquad 
      + 2\, h^{zz} h^{w\bar w}\left({\pi\over 12}\right)
             \partial_z^2 \,\delta (z-w) \\
  &\qquad\qquad\qquad\qquad\qquad 
      + h^{zz} h^{\bar w\bar w}\left(- {\pi\over 12}\right)
      \partial_z \partial_{\bar z} 
       \,\delta (z-w) \\
  &\qquad\qquad\qquad\qquad\qquad + \dots \biggr\} \, \expect{1},
\end{align*}
where we have inserted the contact term two-point functions derived before.
To save space, we only wrote out the first three terms.  Now, using 
$$
 \pi \,\delta(z - w) = \partial_z\partial_{\bar z}\, \ln |z-w|^2, 
$$
and performing partial integrations, we find 
\begin{align*}
&  - {1\over 12} \left({1\over 4 \pi}\right)^2\int d^2z  \int d^2w \,
     \left(-\partial_z^2 h^{zz} -\partial_{\bar z}^2 h^{\bar z \bar z}
     + 2\,\partial_z \partial_{\bar z} h^{z\bar z}\right) \times \\
&\qquad\qquad\qquad\qquad
     \times \ln |z - w|^2  \, \left(-\partial_w^2 h^{ww} -
       \partial_{\bar w}^2 h^{\bar w \bar w}
     + 2\,\partial_w \partial_{\bar w} h^{w\bar w}\right)\\ 
&\qquad =  - {1\over 12} \left({1\over 4 \pi}\right)^2 \int d^2z  \int d^2w\, 
     \delta_g R (z)\,
     \ln |z - w|^2  \, \delta_g R (w).
\end{align*}
Notice that all the $\expect{TT}$ contact terms were necessary
to obtain the correct curvature factors $\delta R$.

Our final result for the plane is\cite{difrancesco,polchinski} 
\begin{align*}
  \delta_g^2
  &\int [d\phi]_g^\mathit{PV} \, e^{-S} 
&= - {1\over 12} \left({1\over 4 \pi}\right)^2 \int d^2z  \int d^2w\, 
     \delta_g R (z)\,
     \ln |z - w|^2  \, \delta_g R (w) \expect{1}.
\end{align*}
Notice that this result is entirely consistent with that of the previous 
section, 
since for a deformation of $g$ by a vector field we have $\delta R = 0$, 
so that the right hand side vanishes.

\section{Conclusion}

In this paper we
presented a coordinate-invariant Pauli-Villars-based
approach to the definition
of the path integral measure 
and the calculation of anomalies in two-dimensional scalar
conformal field theory.    

We showed the agreement, despite seemingly different 
origins, of the conformal anomaly in the Pauli-Villars 
and the Fujikawa approaches. 

The natural, fully regularized 
energy-momentum in our coordinate-invariant approach 
is a true tensor quantity satisfying classical
Ward identities.  
We related this quantity to the more familiar non-tensor 
object arising in the operator formalism.

We provided a direct path-integral 
calculation of the nontrivial contact terms arising
in expectation values of certain energy-momentum products, 
previously derived only using axiomatic considerations.
We used these in a simple consistency check confirming
 the change of
variables formula for the path integral measure.
We also showed that the contact terms are essential to 
obtaining the correct 
relationship between the conformal anomaly and the energy-momentum 
two-point functions in our formalism.   

It is our hope that this work may have some inherent 
interest as an illustration, in a simple model, of the issues
involved in defining a coordinate-invariant path integral
and energy-momentum tensor 
in a matter theory on a nontrivial gravitational background.   

We also hope that this work may be helpful in 
illustrating the origin, often physically 
opaque in the operator formalism, of some simple
anomalous formulas in conformal field theory.
It is important to understand to what extent one can trust
straightforward manipulations of path integrals
to obtain potentially anomalous 
conservation laws and Ward identities.
The conclusion of this paper is that, given a suitably 
coordinate-invariant regularization such as the one defined here, one
can trust these manipulations a great deal.  However, as we have seen, the 
translation of the resulting formulas to other formalisms
may be non-trivial and subtle.

\section*{Acknowledgments}

    I would like to thank Dr. Miquel Dorca for very useful discussions.
    I would also like to thank Prof.\ Antal Jevicki and the Brown
    University Physics department for their support.

\end{document}